\documentclass[12pt,a4paper,superscriptaddress,preprintnumbers,showkeys,pra,Aps]{revtex4-1}
\usepackage[fleqn]{amsmath}
\usepackage{amssymb}
\usepackage{mathrsfs}
\usepackage{bm}
\usepackage{graphicx}
\usepackage{subfigure}
\usepackage{float}
\usepackage{pict2e}
\usepackage{booktabs}
\usepackage{tikz}
\usepackage{ulem} % 
\usepackage[pdfstartview=FitH,
            bookmarksnumbered=true,
            bookmarksopen=true,
            colorlinks=true,
            pdfborder={0 0 1},
            linkcolor=blue,
            citecolor=blue,
            urlcolor=blue
            ]{hyperref}
\usepackage[left=2.5cm, right=2.5cm, top=2.5cm, bottom=2.2cm]{geometry}
\AtBeginDocument{
			\heavyrulewidth=.08em
			\lightrulewidth=.04em
			\cmidrulewidth=.03em
			\belowrulesep=.65ex
			\belowbottomsep=0pt
			\aboverulesep=.4ex
			\abovetopsep=0pt
			\cmidrulesep=\doublerulesep
			\cmidrulekern=.5em
			\defaultaddspace=.5em
			} % set style for TABLE to make {booktabs} compatible with revtex

\begin{document}

\title{Scattering of flexural--gravity waves by a group of elastic plates floating on a stratified fluid}

\author{Q. R. Meng}
\author{D. Q. Lu}
\thanks{Corresponding author. Email: dqlu@shu.edu.cn, dqlu@graduate.hku.hk}
\affiliation{Shanghai Institute of Applied Mathematics and Mechanics, Shanghai University, 149 Yanchang Road, Shanghai 200072, China}
\affiliation{Shanghai Key Laboratory of Mechanics in Energy Engineering, 149 Yanchang Road, Shanghai 200072, China}

\date{\today}

\begin{abstract}

 A hydroelastic problem of flexural--gravity waves scattering by a demarcation between two floating elastic plates is investigated within the frame of linear potential-flow theory, where the method of matched eigenfunction expansions is employed for analysis. A generalized extension is subsequently derived to promote the formulae to the case of multiple elastic plates on a stratified fluid with multiple layers, which is helpful to study the hydrodynamic behaviors of inhomogeneous floating covers as well as the effects of density stratification in seawater. The eigenfunction expansions are numerically calculated by an inner product technique, in which an orthogonal definition involving an explicit differential term exhibits the effectiveness in dealing with the multi-layer matching relations between adjacent regions covered by different elastic plates. By use of Green's theorem, an energy conversation relation is deduced to assure the convergence of the calculation to the physical reality, and the high converging rates are exhibited afterwards. The correctness of numerical results are also verified by comparing with a previous analytical method. The details of the hydrodynamic responses of the generalized extension, {especially the impact of the fluid stratification on the inner forces of the elastic plates,} are discussed under different situations.

\end{abstract}


\maketitle

\section{\label{Sec.1.intro}Introduction}

 It has been long perceived that inhomogeneous profiles commonly exist in the covers floating on seawater, which usually refer to ice floes in polar regions or some artificial architecture called Very Large Floating Structures (VLFSs). For the ice floes, the inhomogeneity is normally embodied in the natural undulant topography, e.g.\ refrozen leads, pressure ridges and cracks (see Squire et al.\ \cite{1995-Squire-p115}). While for the VLFSs, which are basically premised on exploiting expansive oceanic spaces, the inhomogeneity is designed intentionally for {functional and economical requirements and better hydrodynamic performances}. The inhomogeneous properties will lead to nontrivial results when we consider the interaction between the motion of flexural--gravity waves and the deformation of the covers. Substantial literatures have proposed opinions for the modeling of the inhomogeneous covers, where the ice floes are often represented by infinite elastic plates with continuous or stepwise changing profiles (e.g.\ Williams and Squire \cite{2004-Williams-p3469}, Williams and Squire \cite{2006-Williams-p113}), and the VLFSs are generally simplified by multiple finite elastic plates with connections (e.g.\ Xia et al.\ \cite{2000-Xia-p261}, Khabakhpasheva and Korobkin \cite{2002-Khabakhpasheva-p21}). The study on the hydrodynamic responses of the inhomogeneous elastic plate is a significant topic to improve the knowledge of hydroelasticity.

 In this paper, we firstly consider the two-dimensional scattering problem of flexural--gravity waves going across the boundary between two elastic plates, which is one of the commonplaces due to the inhomogeneity. The physical properties of each plate are homogeneous but permitted to be different from each other, as long as the thickness is thin enough to assure that the Kirchhoff--Love plate theory is applicable to approximate the plate deformation. This topic is very similar to the conventional investigations for the crack existing in uniform ice floes. For example, Marchenko \cite{1999-Marchenko-p511} studied this problem via the variational method and then employed asymptotic series to discuss the edge waves subjected to parametric excitations. Evans and Porter \cite{2003-Evans-p143} used both the Green function method and the eigenfunction expansion method to solve this problem, and uniquely they derived two special functions as the fundamental variables, which {can denote} the reflection and transmission coefficients without complete solutions. Porter and Evans \cite{2014-Porter-p533} investigated the trapped modes induced by a finite length crack existing in the center of a three-dimensional infinite elastic plate, and compared the results {by} two analytical approaches. In general, the crack usually occurs in a slowly changing part of an ice floe, thus it is very reasonable to study a uniform elastic plate. For the boundary between two different elastic plates in other cases, slight modifications should be made for the methods employed in the above-mentioned investigations.

 The floating covers with continuous profiles can be regarded as a limiting case of multiple thin plates fixed together by {built-in} connections. The case of multiple plates has ever been studied by Karmakar et al.\ \cite{2009-Karmakar-p1065}. After ignoring the evanescent modes of the flexural--gravity waves, Karmakar et al.\ \cite{2009-Karmakar-p1065} successfully {obtained} the wave reflection and transmission in the far field. Under a short-wavelength assumption, the evanescent modes have no impact on the solution given by Karmakar et al.\ \cite{2009-Karmakar-p1065}, while it is important for the local responses of the elastic plates according to Fox and Squire \cite{1990-Fox-p11629}. Xia et al.\ \cite{2000-Xia-p261} investigated the hydrodynamic behavior of multiple articulated plates, where the connectors were composed of vertical and rotational springs. Squire and Dixon \cite{2001-Squire-p327} considered the wave reflection and transmission across an ice sheet with periodically and randomly spaced multiple cracks, and a set of discrete periods for the cracks were found so that the waves transmit the ice floe perfectly without reflection. Porter and Evans \cite{2006-Porter-p425} extended the concise expressions in Evans and Porter \cite{2003-Evans-p143} for the reflections and transmissions, and subsequently derived an explicit solution for the case of infinite ice sheet with multiple narrow cracks. Kohout et al.\ \cite{2007-Kohout-p649} carried out a complete calculation for obliquely incident waves with the similar physical model. Different from the literatures mentioned above, the method of Kohout et al.\ \cite{2007-Kohout-p649} is applicable to the multiple elastic plates with variable properties.

 The method of matched eigenfunction expansions will be employed in the present work. It is a fundamental method to fit the potential functions along the matching boundaries. An advantage of this method is that the formulations of the system are possible to be promoted to a generalized case of wave scattering by arbitrary numbers of elastic plates. Applying the method of matched eigenfunction expansions to the case of wave scattering by the group of multiple plates, the challenging difficulty is to have the mass flux and pressure well matched for the boundaries between every two plates. Previous investigations by this method are on the wave scattering by one semi-infinite plate. A prior approach is proposed by Fox and Squire \cite{1990-Fox-p11629, 1994-Fox-p185}, who defined an error function method and minimized it by the least square technique to calculate unknowns. By introducing an inner product technique for the same {problem}, Sahoo et al.\ \cite{2001-Sahoo-p3215} {successfully employed} the vertical eigenfunctions derived from the plate-covered region to make an inner product on the matching boundary and {obtained convergent results}. While for the case of multiple plates, both of these methods will be {dysfunctional}. The error function technique \cite{1990-Fox-p11629, 1994-Fox-p185} is insufficient to figure out results of high precision under acceptable efficiency. The inner product technique \cite{2001-Sahoo-p3215} by the vertical eigenfunctions of plate-covered region, according to Kohout et al.\ \cite{2007-Kohout-p649}, will ``lead to an ill-conditioned system of equations,'' {and it is even} impossible to provide a solvable system, where the number of equations will {exceed} the number of unknowns, for the boundaries between every two elastic plates. Kohout et al.\ \cite{2007-Kohout-p649} presented another manner to make an inner product, which is to use the Fourier basis functions to replace the vertical eigenfunctions of Sahoo et al.\ \cite{2001-Sahoo-p3215}. Under this definition, the hydrodynamic behavior of arbitrary numbers of elastic plate is solved. {Meng and Lu \cite{2017-meng-p295} found the vertical eigenfunctions in the free-surface region have eligible performance in making inner products to deal with a matching relation between a region covered by an elastic plate and one with a semi-immersed floating body.} For our calculation, {with the consideration} that the flexural--gravity wave is a variation from conventional free-surface one, we let the physical properties of the plate tend to zero to obtain a set of vertical eigenfunctions of free-surface waves, by which an inner product is {defined}. An orthogonal relation for this manner is subsequently derived by adding an explicit differential term for the inner product, which {exhibits great improvements} for the calculation efficiency.

 In our problem, the fluid is assumed to be stratified by density, which is a refined approximation to the actual seawater. The fluid stratification has been studied by many authors, for example, Cadby and Linton \cite{2000-Cadby-p155}, Bhattacharjee and Sahoo \cite{2008-Bhattacharjee-p133}. In Sec.\ \ref{Sec.2.formu}, we formulate wave scattering by two elastic plates floating on a 2-layer fluid. Afterwards, in Sec.\ \ref{Sec.3.ext}, a generalized model of multiple elastic plates floating on the stratified layer with multiple layers is also considered. The generalized model can be a universal representation for a class of problems. The orthogonal inner product for the multi-layer fluid with a concise and explicit term provides the possibility for fast calculation. In Sec.\ \ref{Sec.4.Calcu}, several numerical results and validations are exhibited. The rate of convergence and the correctness of the numerical results are verified by Green's theorem and by a comparison with {the method in Kohout et al.\ \cite{2007-Kohout-p649}.} The different structures are discussed for single-plate, 2-plate and 4-plate models. The stratification effect is discussed, although it is usually regarded insignificant in the research about the hydrodynamic responses of structures. After picking several points on a parabolic curve as the density for 4-layer and 8-layer fluid, which is assumed as the real density distribution of the seawater, we found the stratification will have great impact on the shear force of the elastic plate, especially for the middle area of every single plate and the connections. That is a distinct improvement from previous researches.

\section{\label{Sec.2.formu}Formulation of two semi-infinite plate floating on a 2-layer fluid}

\subsection{\label{subSec.2.1.assCon}Assumptions and conditions}

\begin{figure}[ht]
    	\centering
    	\setlength{\unitlength}{0.85cm}
    	\begin{picture}(14,6)
				\linethickness{0.05\unitlength}
				\multiput(1,4)(0,0.2){2}{\line(1,0){12}}
				\put(7,4){\line(0,1){0,2}}
				\linethickness{0.02\unitlength}
				\put(3.4,4.6){\vector(1,0){1.8}}
				\put(3,5){Incident waves}
	   			\put(7,4.7){\vector(1,0){1}}
				\put(7,4.7){\vector(0,1){1}}
				\put(8.1,4.7){$x$}
				\put(7.1,5.7){$z$}
				\put(14,3.9){$z=0$}
				\put(14,2.5){$z=-h_1$}
				\put(14,-0.1){$z=-H$}
				\put(-1.2,1.2){Layer 2}
				\put(-1.2,3.2){Layer 1}
				\multiput(1,2.6)(0.2,0){60}{\line(1,0){0.1}}
				\multiput(7,0)(0,0.2){20}{\line(0,1){0.1}}
				\multiput(1.1,0)(0.1,0){120}{\line(-1,-1){0.1}}
				\put(1,0){\line(1,0){12}}
		\end{picture}
	    \caption{{Incident flexural--gravity waves} across the boundary between the domains covered by diverse elastic plates}
	    \vspace{0mm}
		\label{Fig.schDia}
	\end{figure}
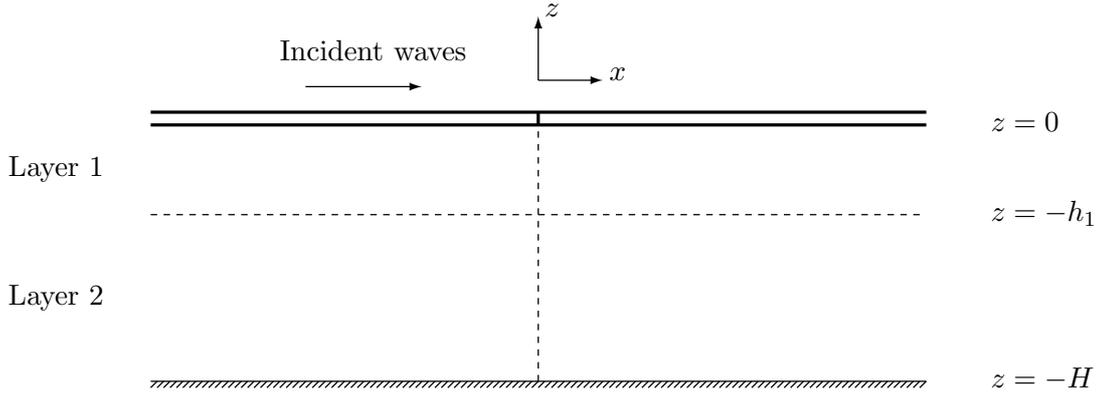

 {Waves traveling under the regime of thin elastic plates, i.e.\ flexural--gravity waves, usually exhibit different physical attributes because of the synthetic action of the elastic restoring forces and gravitational forces.}  We consider a simple situation as shown in Fig.\ \ref{Fig.schDia} to demonstrate the scattering problem of flexural--gravity waves. A 2-layer fluid is entirely covered by two semi-infinite elastic plates, where the fluid is assumed to be ideal and incompressible, the motion {to be} irrotational, and the elastic plates {to be} homogeneous and isotropic. {{Waves traveling under the regime of elastic plates, i.e.\ flexural--gravity waves, are commonly caused by the synthetic action of the elastic restoring and gravitational forces.}} The densities and the thicknesses of the stratified fluid are denoted by $\rho_m$ and $h_m$ respectively, with $m=1$ for the upper layer and $m=2$ for the lower one ($\rho_1 < \rho_2$). The physical properties of the elastic plates are permitted to be diverse, thus we introduce $D_n$ and $M_n$ to denote the flexural rigidity and the mass per unit length respectively, with $n=1$ for the left plate and $n=2$ for the right one. Given that the immersed depths of {thin plates are usually shallow enough to be neglected, we assume the plates adhere to the water surface closely with no cavitation.} {A right-handed coordinate} system is located at the intersecting point of the two plates, {then} the positions of interface and bottom are dimensioned {as} $z=-h_1$ and $z=-H$, where $H=h_1+h_2$.

 {The linear potential-flow theory is employed hereinafter.} Focusing on a specific frequency $\omega$, the motion of the flexural--gravity waves can be described by a velocity potential function $\phi(x,z,t) = \Re [\Phi(x,z)\textup{e}^{-\textup{i} \omega t}]$, where $\Phi(x,z)$ is a complex-valued function with respect to the spatial variables only. In the whole fluid domain, $\Phi(x,z)$ obeys the governing equation
	\begin{align}
			\label{Eq.govEq}
	    	&\nabla^2 \Phi(x,z) = 0.
	\end{align}
 The surface displacement {of the elastic plates} can also be quantified by a similar form, $\Re [\zeta(x) \textup{e}^{-\textup{i} \omega t}]$. On the surface, the interface and the bottom of the fluid, the potential function satisfies the kinematic conditions as follows:
 	\begin{align}
			\label{Eq.bonCon1}
			&\frac{\partial \Phi}{\partial z} \Big|_{z=0}= -\textup{i} \omega \zeta, &&(-\infty<x<+\infty),
	\end{align}
	\begin{align}
			\label{Eq.bonCon2}
			&\frac{\partial \Phi}{\partial z} \Big|_{z=-h_1^+} = \frac{\partial \Phi}{\partial z} \Big|_{z=-h_1^-},&&(-\infty<x<+\infty),
	\end{align}
	\begin{align}
			\label{Eq.bonCon3}
			&\frac{\partial \Phi}{\partial z} \Big|_{z=-H} = 0,&&(-\infty<x<+\infty).
	\end{align}
 Considering the {force balance} on the plate-fluid contact plane ($z=0$) and the continuity of the pressure on the interface ($z=-h_1$), we can derive the dynamic conditions:
	\begin{align}
			\label{Eq.bonCon4}
			& \rho_1 \omega^2 \Phi - \left( D_n \frac{\partial^4}{\partial x^4} - M_n \omega^2 + \rho_1 g \right) \frac{\partial \Phi}{\partial z} = 0, && (-\infty< x < +\infty,z=0),
	\end{align}
	\begin{align}
			\label{Eq.bonCon5}
			&\gamma \left[ K \Phi - \frac{\partial \Phi}{\partial z} \right]_{z=-h_1^+} = \left[ K \Phi - \frac{\partial \Phi}{\partial z}\right]_{z=-h_1^-}, && (-\infty<x<+\infty),
	\end{align}
 where $g$ is the gravitational acceleration, $\gamma = \rho_1/\rho_2$ and $K = \omega^2/g$. Depending on what connecting situation is specified at $(0,0)$, {four connection conditions related to the deflections, rotational angles, bending moments, and shear forces of the plates can be found}. A unified representation for various situations can be denoted in a consolidated equation of
	\begin{align}
			 \label{Eq.edgCon}
			 \bm{E}^- \bm{\lambda}^- = \bm{E}^+ \bm{\lambda}^+,
	\end{align}
 where $\bm{\lambda}^\pm = \left[ \zeta(0^\pm), \zeta'(0^\pm), \zeta''(0^\pm), \zeta'''(0^\pm) \right]^\intercal$ and $\bm{E}^\pm$ are two $4$ by $4$ matrixes. For a given connecting type, $\bm{E}^{\pm}$ can be obtained readily.

\subsection{\label{subSec.2.2.sol}Method of solution}

 Substituting the general solution of Laplace's equation into Eqs.\ (\ref{Eq.bonCon2}),  (\ref{Eq.bonCon3}) and (\ref{Eq.bonCon5}), we can obtain a piecewise vertical eigenfunction for the 2-layer fluid:
	\begin{align}
			\label{Eq.verEigen}
			V(k,z) = \dfrac{ K \cosh k h_2 - \varepsilon k \sinh k h_2 }{\gamma K \cosh k H} \cosh k(z+h_1) + \dfrac{\sinh k h_2 }{\cosh k H} \sinh k(z+h_1),
	\end{align}
 for $-h_1<z<0$ while
	\begin{align}
			V(k,z) = \dfrac{\cosh k(z+H)}{\cosh k H},
	\end{align}
 for $-H<z<-h_1$, where $k$ refers to the wave number and $\varepsilon = 1 - \gamma$. {By using the surface boundary condition Eq.\ (\ref{Eq.bonCon4}),} the dispersion relation can be deduced as follows:
	\begin{align}
			\label{Eq.disRel}
			\left( t_0 G_n + \gamma t_1 t_2 + 1 \right) \omega^4 - \left(  t_0 F_n + \varepsilon t_2 + \varepsilon t_1 t_2 G_n \right) \omega_0^2 \omega^2 + \varepsilon F_n t_1 t_2 \omega_0^4 = 0,
	\end{align}
 where $\omega_0^2= g k$, $t_1=\tanh k h_1$, $t_2 = \tanh k h_2$, $t_0 = t_1 + \gamma t_2$, $G_n = M_n k / \rho_1$, and $F_n = D_n k^4  / \rho_1 g + 1$, with $n=1,2$ for different plate-covered regions.

 For a given $\omega$, we can {numerically calculate the wave numbers for each elastic plate region from Eq.\ (\ref{Eq.disRel})}. The flexural--gravity waves in a 2-layer fluid have four real roots $\pm \tilde{k}_{n,0_m}$ ($m=1,2$ for surface and interfacial wave modes), two couples of conjugate complex roots $\pm \textup{i} \tilde{k}_{n,j}$ ($j=\textup{I},\textup{II}$) and infinite numbers of pure imaginary roots $\pm \textup{i} \tilde{k}_{n,j}$ ($j=1,2,\cdots$), where the first subscript $n$ is employed to discern the different regions and the second one is to denote the sequence of wave modes with $0_m$ for propagating waves, $j=\textup{I},\textup{II}$ for attenuated or augmented propagating waves and $j=1,2,\cdots$ for evanescent waves.

 In the region $x<0$, the potential function consists of two parts, namely incident potential $\Phi_{\textrm{I}_1}(x,z)$ and reflection potential $\Phi_{\textrm{R}_1}(x,z)$, while in $x>0$, the potential function only contains the transmission potential $\Phi_{\textrm{T}_2}(x,z)$:
 	\begin{align}
	 	\label{Eq.potDec}
	 	&\Phi(x,z) = \left\{
	 				 \begin{array}{ll}
	 					\Phi_{\textrm{I}_1}(x,z) + \Phi_{\textrm{R}_1}(x,z), & (x<0),\\
	 					\Phi_{\textrm{T}_2}(x,z), & (x>0),
	 				 \end{array}
	 				 \right.
 	\end{align}
{where} the subscripts ``1" and ``2" denote different plate-covered regions.
 The incident potential should be prescribed as {a} known constituent to the system, and the reflection and transmission potentials can be expanded according to the wave numbers and far-field conditions,
	\begin{align}
	 	\label{Eq.potExpR}
	 	&\Phi_{\textrm{R}_1}(x,z) = \sum_m R_{1,0_m} \textup{e}^{- \textup{i} \tilde{k}_{1,0_m} x} \tilde{Z}_{1,0_m} + \sum_j R_{1,j} \textup{e}^{\tilde{k}_{1,j} x } \tilde{Z}_{1,j},
	\end{align}
	\begin{align}
	 	\label{Eq.potExpT}
	 	&\Phi_{\textrm{T}_2}(x,z) = \sum_m T_{2,0_m} \textup{e}^{\textup{i} \tilde{k}_{2,0_m} x } \tilde{Z}_{2,0_m} + \sum_j T_{2,j} \textup{e}^{- \tilde{k}_{2,j} x } \tilde{Z}_{2,j},
	\end{align}
 where
	\begin{align}
			\label{Eq.verdef}
			\left\{ \tilde{Z}_{n,0_m}, \tilde{Z}_{n,j} \right\} = \left\{ V( \tilde{k}_{n,0_m},z), V( \textup{i} \tilde{k}_{n,j},z) \right\}, && (n=1,2).
	\end{align}
 $\sum_m$ and $\sum_j$ are the summation sign of $m=1,2$, $j=\textup{I},\textup{II},1,2,\cdots$, respectively. $R_{1,0_m}$, $R_{1,j}$ and $T_{2,0_m}$, $T_{2,j}$ are respectively the {unknown coefficients corresponding to} the reflection and transmission modes. The potential {functions satisfy} the matching conditions for the continuities of pressure and of velocity along $x=0$, which lead to
	\begin{align}
			\label{Eq.mthCon}
			&\Phi_{\textrm{I}_1} + \Phi_{\textrm{R}_1} = \Phi_{\textrm{T}_2}, & &\dfrac{\partial (\Phi_{\textrm{I}_1} + \Phi_{\textrm{R}_1})}{\partial x} = \dfrac{\partial \Phi_{\textrm{T}_2}}{\partial x}, & & (x = 0,-H < z < 0).
	\end{align}

 In order to calculate the reflection and transmission coefficients, one appropriate method should be applied to deal with Eq.\ (\ref{Eq.mthCon}). Numbers of previous literatures have proposed various techniques to calculate the expansions. For example, Fox and Squire \cite{1990-Fox-p11629, 1994-Fox-p185} tried to define an error function and used the least square method to compute the coefficients in the case of free-surface waves incoming to a semi-infinite plate. Sahoo et al.\ \cite{2001-Sahoo-p3215} and Xu and Lu \cite{2010-Xu-p809} adopted inner product methods to calculate the same problem, where the vertical eigenfunction of flexural--gravity waves and free-surface gravity waves were employed for definition, respectively. Khabakhpasheva and Korobkin \cite{2002-Khabakhpasheva-p21} and Kohout et al.\ \cite{2007-Kohout-p649} mentioned another manner of making an inner product, in which they replaced the vertical eigenfunctions of Sahoo et al.\ \cite{2001-Sahoo-p3215} and Xu and Lu \cite{2010-Xu-p809} by the Fourier basis functions, and obtained convergent results. 

 For the problem in this section, between both sides of the matching boundary are all flexural--gravity waves. If we employ the vertical eigenfunctions on either side to make an inner product, it will ``lead to an ill-conditioned system of equations'' (Kohout et al.\ \cite{2007-Kohout-p649}). Considering that the free-surface waves can be thought of as a limiting case from the flexural--gravity waves, we attempt to employ the vertical eigenfunctions of free-surface waves to define the inner product. This set of vertical eigenfunctions can be derived by artificially degenerating the plate-covered region to a free-surface one after having $D_n$ and $M_n$ tend to zero. The dispersion relation for the free-surface waves is
 	\begin{align}
			\label{Eq.disRelfree}
			( \gamma t_1 t_2 + 1 ) \omega^4 - ( t_1 + t_2 ) \omega_0^2 \omega^2 + \varepsilon t_1 t_2 \omega_0^4 = 0.
	\end{align}
 For the same $\omega$, we can find four real roots $\pm k_{0_m}$ ($m=1,2$) and infinite numbers of pure imaginary roots $\pm \textup{i} k_i$ ($i=1,2,\cdots$) for Eq.\ (\ref{Eq.disRelfree}). The two couples of conjugate complex roots for flexural--gravity waves disappear. We write the vertical eigenfunctions of free-surface waves in the form of
 	\begin{align}
 			\label{Eq.verdef2}
			\left\{ Z_{0_m}, Z_i \right\} = \left\{ V(k_{0_m},z), V(\textup{i} k_i,z) \right\}.
	\end{align}
 
 The inner product is defined for two vertical eigenfunctions, e.g.\ $U(z)$ and $V(z)$, as follows:
	\begin{align}
			\label{Eq.inPdef}
			\left\langle U,V \right\rangle = \int_{-H}^{-h_1} U \cdot V \,\textup{d}z + \gamma \int_{-h_1}^0 U \cdot V \,\textup{d}z.
	\end{align}
 We employ the vertical eigenfunctions $Z_p$ to make an inner product for Eqs.\ (\ref{Eq.mthCon}) by the sequence of $p = 0_1,0_2,1,2,\cdots$, {and subsequently} two sets of equations are obtained as follows:
 	\begin{align}
		 	\label{Eq.mthEq1}
		 	&\left\langle \Phi_{\textrm{R}_1}, Z_p \right\rangle - \left\langle \Phi_{\textrm{T}_2}, Z_p \right\rangle = - \left\langle \Phi_{\textrm{I}_1}, Z_p \right\rangle,
	\end{align}
	\begin{align}
		 	\label{Eq.mthEq2}
		 	&\left\langle \frac{\partial \Phi_{\textrm{R}_1}}{\partial x}, Z_p \right\rangle - \left\langle \frac{\partial \Phi_{\textrm{T}_2}}{\partial x}, Z_p \right\rangle = - \left\langle \frac{\partial \Phi_{\textrm{I}_1}}{\partial x}, Z_p \right\rangle.
	\end{align}
 For the two sort of vertical eigenfunctions, we can derive an orthogonal relation by adding an explicit differential term, namely
 	\begin{align}
		\label{Eq.inPorth}
			&\left\langle \tilde{Z}_{n,q}, Z_p \right\rangle - \mathscr{D}_n(p,q) = 0, &&
     (p = 0_1,0_2,1,2,\cdots;q = 0_1,0_2,\textup{I},\textup{II},1,2,\cdots),
	\end{align}
 where
 	\begin{align}
			\mathscr{D}_n(p,q) = \dfrac{ (D_n \tilde{k}^4_{n,q} - M_n \omega^2)}{\rho_2 \omega^2 (k^4_p - \tilde{k}^4_{n,q})} \left[ \dfrac{\partial \tilde{Z}_{n,q}}{\partial z}\frac{\partial^3 Z_p}{\partial z^3} + \dfrac{\partial^3 \tilde{Z}_{n,q}}{\partial z^3} \frac{\partial Z_p}{\partial z}  \right]_{z=0}, &&(n=1,2).
	\end{align}
 Conspicuously, the calculation will be improved by this explicit expression for the inner product. We truncate the potential functions of Eqs.\ (\ref{Eq.potExpR}) and (\ref{Eq.potExpT}) at $j=S$, and then with the help of Eq.\ (\ref{Eq.inPorth}), Eqs.\ (\ref{Eq.mthEq1}) and (\ref{Eq.mthEq2}) can be rewritten {in the form} of a partitioned matrix equation, as follows:
 	\begin{align}
			\label{Eq.mthMat}
			\left(
			\begin{array}{cc}
					\bm{M}_{\textrm{R}_1}  & \bm{M}_{\textrm{T}_2} \\
					\bm{N}_{\textrm{R}_1} & \bm{N}_{\textrm{T}_2}
			\end{array}
			\right)
			\left(
			\begin{array}{cc}
					\bm{\alpha}_{\textrm{R}_1} \\
					\bm{\alpha}_{\textrm{T}_2}
			\end{array}
			\right)
			=
			\left(
			\begin{array}{cc}
					\bm{\beta} \\
					\bm{\beta}_x
			\end{array}
			\right).
		\end{align}
 where $\bm{M}_{\textrm{R}_1}$, $\bm{M}_{\textrm{T}_2}$, $\bm{N}_{\textrm{R}_1}$, and $\bm{N}_{\textrm{T}_2}$ are $S+2$ by $S+4$ matrixes. Every element in these matrixes is related to the inner product by relevant wave numbers, thus we use $(p,q)$ with $p=0_1,0_2,1,\cdots,S$ and $q=0_1,0_2,\textup{I},\textup{II},1,\cdots,S$ as the representative of corresponding row and column numbers. The details for the matrixes are
 	\begin{align}
	 		&\bm{M}_{\textrm{R}_1}(p,q) = \mathscr{D}_1 (p,q), & &\bm{M}_{\textrm{T}_2}(p,q) = - \mathscr{D}_2 (p,q),
	\end{align}
 	\begin{align}
		&\bm{N}_{\textrm{R}_1}(p,q) = \delta \tilde{k}_{1,q}\mathscr{D}_1(p,q), & &\bm{N}_{\textrm{T}_2}(p,q) = \delta \tilde{k}_{2,p} \mathscr{D}_2(p,q),
	\end{align}
 where $\delta = 1$ for $q=0_1,0_2$ while $\delta = \textup{i}$ for $q=\textup{I},\textup{II},1,\cdots,S$. $\bm{\beta}$ and $\bm{\beta}_x$ are $(S+2)$-dimensional column vectors, which can be determined in terms of the incident potential:
 	\begin{align}
			 	&\bm{\beta}(p) = - \left\langle \Phi_{\textrm{I}_1} , Z_p \right\rangle, & & \bm{\beta}_x(p) = - \left\langle \dfrac{\partial \Phi_{\textrm{I}_1}}{\partial x} , Z_p \right\rangle.
 	\end{align}
  $\bm{\alpha}_{\textrm{R}_1}$ and $\bm{\alpha}_{\textrm{T}_2}$ are $(S+4)$-dimensional column vectors related to the coefficients to be determined:
 	\begin{align}
 				\bm{\alpha}_{\textrm{R}_1} &= \left[ R_{1,0_1},R_{1,0_2},R_{1,\textup{I}},R_{1,\textup{II}},R_{1,1},\cdots,R_{1,S} \right]^\intercal,
 	\end{align}
 	\begin{align}
 				\bm{\alpha}_{\textrm{T}_2} &= \left[ T_{2,0_1},T_{2,0_2},T_{2,\textup{I}},T_{2,\textup{II}},T_{2,1},\cdots,T_{2,S} \right]^\intercal.
 	\end{align}

 {By means of the relation between the potential function and the surface displacement} in Eq.\ (\ref{Eq.bonCon1}), four additional equations can be obtained from Eq.\ (\ref{Eq.edgCon}) as well. {Associating with }Eq.\ (\ref{Eq.mthMat}), a simultaneous system of $2S+8$ equations for $2S+8$ unknown coefficients is established. The spatial potential functions can be calculated numerically.

\section{\label{Sec.3.ext}Generalized extension to multiple elastic plates on a stratified fluid with multiple layers}

\subsection{\label{subSec.3.1.potExp}Potential functions}

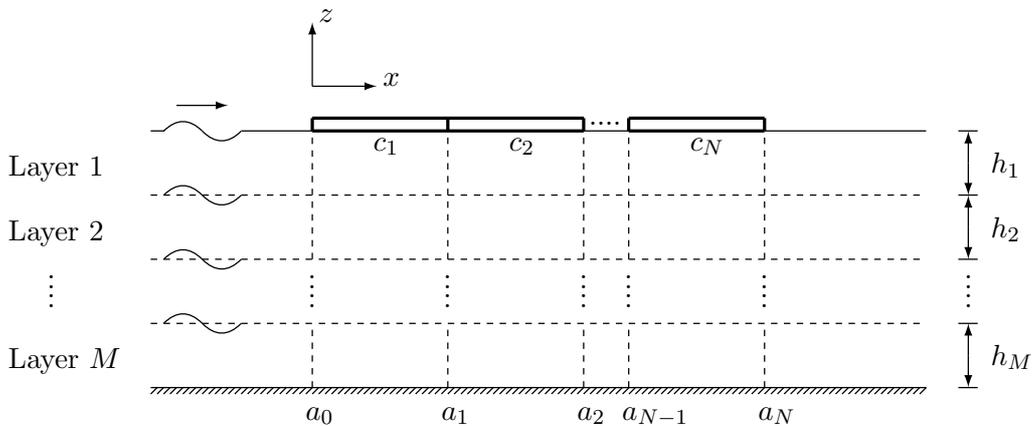
\begin{figure}[ht]
	    \centering
	    \setlength{\unitlength}{0.85cm}
	    \begin{picture}(14,6)
				\linethickness{0.05\unitlength}
				\multiput(3.5,4)(0,0.2){2}{\line(1,0){4.2}}
				\multiput(8.4,4)(0,0.2){2}{\line(1,0){2.1}}
				\put(7.78,4.1){$....$}
				\put(3.5,4){\line(0,1){0.2}}
				\put(5.6,4){\line(0,1){0.2}}
				\put(7.7,4){\line(0,1){0.2}}
				\put(8.4,4){\line(0,1){0.2}}
				\put(10.5,4){\line(0,1){0.2}}
				\linethickness{0.02\unitlength}
				\put(4.45,3.7){$c_1$}
				\put(6.55,3.7){$c_2$}
				\put(9.35,3.7){$c_N$}
				\put(1,4){\line(1,0){0.21}}
				\put(2.4,4){\line(1,0){1.1}}
				\put(7.7,4){\line(1,0){0.8}}
				\put(10.5,4){\line(1,0){2.5}}
				\qbezier(1.2,4)(1.5,4.3)(1.8,4)
				\qbezier(1.8,4)(2.1,3.7)(2.4,4)
				\put(1.4,4.4){\vector(1,0){0.8}}
				\qbezier(1.2,3)(1.5,3.3)(1.8,3)
				\qbezier(1.8,3)(2.1,2.7)(2.4,3)
				\qbezier(1.2,2)(1.5,2.3)(1.8,2)
				\qbezier(1.8,2)(2.1,1.7)(2.4,2)
				\qbezier(1.2,1)(1.5,1.3)(1.8,1)
				\qbezier(1.8,1)(2.1,0.7)(2.4,1)
	   			\put(3.5,4.7){\vector(1,0){1}}
				\put(3.5,4.7){\vector(0,1){1}}
				\put(4.6,4.7){$x$}
				\put(3.6,5.7){$z$}
				\put(-1.2,3.3){Layer 1}
				\put(-1.2,2.3){Layer 2}
				\put(-1.2,0.3){Layer $M$}
				\put(14,3.35){$h_1$}
				\put(14,2.35){$h_2$}
				\put(14,0.35){$h_M$}
				\put(3.4,-0.5){$a_0$}
				\put(5.5,-0.5){$a_1$}
				\put(7.6,-0.5){$a_2$}
				\put(8.3,-0.5){$a_{N-1}$}
				\put(10.4,-0.5){$a_N$}
				\multiput(13.5,0)(0,1){5}{\line(1,0){0.3}}
				\multiput(13.65,2.5)(0,1){2}{\vector(0,-1){0.5}}
				\multiput(13.65,2.5)(0,1){2}{\vector(0,1){0.5}}
				\put(13.65,0.5){\vector(0,-1){0.5}}
				\put(13.65,0.5){\vector(0,1){0.5}}
				\multiput(1,3)(0.2,0){60}{\line(1,0){0.1}}
				\multiput(1,2)(0.2,0){60}{\line(1,0){0.1}}
				\multiput(1,1)(0.2,0){60}{\line(1,0){0.1}}
				\multiput(3.5,0)(0,0.2){5}{\line(0,1){0.1}}
				\multiput(5.6,0)(0,0.2){5}{\line(0,1){0.1}}
				\multiput(7.7,0)(0,0.2){5}{\line(0,1){0.1}}
				\multiput(8.4,0)(0,0.2){5}{\line(0,1){0.1}}
				\multiput(10.5,0)(0,0.2){5}{\line(0,1){0.1}}
				\multiput(3.5,2)(0,0.2){10}{\line(0,1){0.1}}
				\multiput(5.6,2)(0,0.2){10}{\line(0,1){0.1}}
				\multiput(7.7,2)(0,0.2){10}{\line(0,1){0.1}}
				\multiput(8.4,2)(0,0.2){10}{\line(0,1){0.1}}
				\multiput(10.5,2)(0,0.2){10}{\line(0,1){0.1}}
				\multiput(-0.6,1.25)(0,0.15){4}{$.$}
				\multiput(3.435,1.25)(0,0.15){4}{$.$}
				\multiput(5.535,1.25)(0,0.15){4}{$.$}
				\multiput(7.635,1.25)(0,0.15){4}{$.$}
				\multiput(8.335,1.25)(0,0.15){4}{$.$}
				\multiput(10.435,1.25)(0,0.15){4}{$.$}
				\multiput(13.59,1.25)(0,0.15){4}{$.$}
				\multiput(1.1,0)(0.1,0){120}{\line(-1,-1){0.1}}
				\put(1,0){\line(1,0){12}}
		\end{picture}
	    \caption{Flexural--gravity wave scattering by multiple elastic plates floating on a stratified fluid with multiple layers}
	    \vspace{0mm}
		\label{Fig.schDiaMulti}
	\end{figure}

 We consider a generalized situation that $N$ finite elastic plates with variable properties floating on a $M$-layer fluid, which can be seen as a multi-module VLFS on the stratified ocean. The subscripts $n=1,2,\cdots,N$ and $m=1,2,\cdots,M$ are applied to mark each single plate and fluid layer, respectively. The elastic plates are continuously placed on the right side of the $z$-axis, as shown in Fig.\ \ref{Fig.schDiaMulti}. The length and midpoint of the $n$-th plate are assigned as $2 L_n$ and $(c_n,0)$. The flexural rigidity and the mass per unit length keep the same symbols, i.e.\ $D_n$ and $M_n$, as in Sec.\ \ref{Sec.2.formu}. The positions of every matching boundary from left to right are denoted by $x = a_0,a_1,a_2,\cdots,a_{N-1},a_N$, where $a_0 = 0$. The density and thickness for the $m$-th layer is given by $\rho_m$ and $h_m$, and then every interface as well as the seabed can be located at $z = - H_m = - (h_1+ \cdots +h_m) $.

 The vertical eigenfunction $V(k,z)$ is also a piecewise one, each component of which has the form of
 	\begin{align}
		 	\label{Eq.verEigenExt}
			V(k,z) = A_m \cosh k (z+H_m) + B_m \sinh k (z+H_m), && (-H_m<z<-H_{m-1}),
	\end{align}
 where $A_m$ and $B_m$ are the coefficients that can be derived by $A_M$ and $B_M$ for the $M$-th layer and the recursion formula as follows:
 	\begin{align}
		 	&\left(
		 	\begin{array}{cc}
				 	A_M \\
				 	B_M
		 	\end{array}
		 	\right)
		 	=
		 	\left(
		 	\begin{array}{cc}
				 	\dfrac{1}{\cosh k H_M} \\
				 	0
		 	\end{array}
		 	\right),
 	\end{align}
 	\begin{align}
 			\label{Eq.verIter}
		 	&\left(
		 	\begin{array}{cc}
		 			A_m \\
		 			B_m
		 	\end{array}
		 	\right)
		 	=
		 	\cosh k h_{m+1}
		 	\left(
		 	\begin{array}{cc}
			\dfrac{1}{\gamma_m} - \dfrac{\varepsilon_m k t_{m+1}}{\gamma_m K} & \dfrac{t_{m+1}}{\gamma_m} - \dfrac{\varepsilon_m k}{\gamma_m K} \\
				 	t_{m+1} & 1
		 	\end{array}
		 	\right)
		 	\left(
		 	\begin{array}{cc}
		 			A_{m+1} \\
		 			B_{m+1}
		 	\end{array}
		 	\right),
	\end{align}
 with $m=1,\cdots,M-1$, $t_m = \tanh k h_m$, $\gamma_m = \rho_m / \rho_{m+1}$, and $\varepsilon_m = 1 - \gamma_m$. The dispersion relation of the region underneath the $n$-th plate is denoted by
	\begin{align}
			\label{Eq.disperIter}
			\left(
			\begin{array}{cc}
					t_1 C_n - K, C_n - K t_1
			\end{array}
			\right)
			\prod_m
		 	\left(
		 	\begin{array}{cc}
				 	\dfrac{1}{\gamma_m} - \dfrac{\varepsilon_m k t_{m+1}}{\gamma_m K} & \dfrac{t_{m+1}}{\gamma_m} - \dfrac{\varepsilon_m k}{\gamma_m K} \\
				 	t_{m+1} & 1
		 	\end{array}
		 	\right)
		 	\left(
		 	\begin{array}{cc}
				 	1 \\
				 	0
		 	\end{array}
		 	\right)
			= 0,
	\end{align}
 where $C_n = k F_n - K G_n $; $\prod_m$ is the multiplication sign of $m=1,2,\cdots,M-1$ for matrix. The definitions of $F_n$ and $G_n$ are consistent with those in Eq.\ \ref{Eq.disRel}. When $D_n$ and $M_n$ tend to zero for a limiting case, we can also obtain the dispersion relation of the free-surface waves in the $M$-layer fluid. For a given $\omega$, the wave numbers can be sought out empirically. For Eq.\ (\ref{Eq.disperIter}), we can find $2M$ real roots $\pm \tilde{k}_{n,0_m}$ ($m=1,2,\cdots,M$), two couples of complex conjugates $\pm \textup{i} \tilde{k}_{n,j}$ ($j=\textup{I},\textup{II}$) and infinite numbers of pure imaginary roots $\pm \textup{i} \tilde{k}_{n,j}$ ($j=1,2,\cdots$), while for the limiting case, we can find $2M$ real roots $\pm k_{0_m}$ ($m=1,2,\cdots,M$) and infinite numbers of pure imaginary roots $\pm \textup{i} k_i$ ($i=1,2,\cdots$).

 The potential function should be separated as
	\begin{align}
		 	\label{Eq.potDecExt}
		 	&\Phi(x,z) = \left\{
		 				 \begin{array}{ll}
		 					\Phi_{\textrm{I}_0}(x,z) + \Phi_{\textrm{R}_0}(x,z), & (x<0),\\
		 					\Phi_{\textrm{T}_n}(x,z) + \Phi_{\textrm{R}_n}(x,z), & (a_{n-1}<x<a_n,~~n=1,\cdots,N),\\
		 					\Phi_{\textrm{T}_{N+1}}(x,z), & (x>a_N).
		 				 \end{array}
		 				 \right.
 	\end{align}
 where the components are expanded in the series of
	\begin{align}
		 	\label{Eq.potExpIExt}
		 	&\Phi_{\textrm{I}_0}(x,z) = \sum_m I_{0,0_m} \textup{e}^{\textup{i} k_{0_m} x} Z_{0_m},
	\end{align}
	\begin{align}
		 	\label{Eq.potExpRExt}
		 	&\Phi_{\textrm{R}_0}(x,z) = \sum_m R_{0,0_m} \textup{e}^{- \textup{i} k_{0_m} x} Z_{0_m} + \sum_i R_{0,i} \textup{e}^{ k_i x } Z_i,
	\end{align}
	\begin{align}
		 	\label{Eq.potExpTnExt}
		 	&\Phi_{\textrm{T}_n}(x,z) = \sum_m T_{n,0_m} \textup{e}^{\textup{i} \tilde{k}_{n,0_m} (x - c_n) } \tilde{Z}_{n,0_m} + \sum_j T_{n,j} \textup{e}^{- \tilde{k}_{n,j} (x - c_n) } \tilde{Z}_{n,j},
	\end{align}
	\begin{align}
		 	\label{Eq.potExpRnExt}
		 	&\Phi_{\textrm{R}_n}(x,z) = \sum_m R_{n,0_m} \textup{e}^{- \textup{i} \tilde{k}_{n,0_m} (x - c_n) } \tilde{Z}_{n,0_m} + \sum_j R_{n,j} \textup{e}^{\tilde{k}_{n,j} (x - c_n) } \tilde{Z}_{n,j} ,
	\end{align}
	\begin{align}
		 	\label{Eq.potExpTExt}
		 	&\Phi_{\textrm{T}_{N+1}}(x,z) = \sum_m T_{N+1,0_m} \textup{e}^{\textup{i} k_{0_m} (x - a_N)} Z_{0_m} + \sum_i T_{N+1,i} \textup{e}^{- k_i (x - a_N)} Z_i,
	\end{align}
 with
	\begin{align}
			&I_{0,0_m} = -\textup{i} \omega \xi_m \left[ \frac{\partial V(k_{0_m},-H_{m-1})}{\partial z}\right]^{-1}.
	\end{align}
 The definitions of $Z_{0_m}$, $Z_i$, $\tilde{Z}_{n,0_m}$, and $\tilde{Z}_{n,j}$ keep the same with Eqs.\ (\ref{Eq.verdef}) and (\ref{Eq.verdef2}).

\subsection{\label{subSec.3.2.matEq}Matching equations}

 The definition of the inner product for the $M$-layer fluid is defined by
 	\begin{align}
			\label{Eq.inPdefS}
			\left\langle U,V \right\rangle = \sum_{m=1}^M \frac{\rho_{m}}{\rho_M} \int_{-H_m}^{-H_{m-1}} U \cdot V \,\textup{d}z,
	\end{align}
 {where we have $H_0 = 0$ for $m=1$}. After altering the differential term to
 	\begin{align}
			&\mathscr{D}_n(p,q) = \frac{ (D_n \tilde{k}^4_{n,q} - M_n \omega^2)}{\rho_M \omega^2 (k^4_p - \tilde{k}^4_{n,q})} \left[ \frac{\partial \tilde{Z}_{n,q}}{\partial z}\frac{\partial^3 Z_p}{\partial z^3} + \frac{\partial^3 \tilde{Z}_{n,q}}{\partial z^3} \frac{\partial Z_p}{\partial z}  \right]_{z=0},\\
			&\quad (p=0_1,0_2,\cdots,0_M,1,2,\cdots;q=0_1,0_2,\cdots,0_M,\textup{I},\textup{II},1,2,\cdots),
	\end{align}
 the orthogonal relation presented in Eq.\ (\ref{Eq.inPorth}) is still practicable for the multi-layer model. Truncating the series at $j=S$ and taking the vertical eigenfunctions of free-surface waves $Z_p$ ($p=0_1,0_2,\cdots,0_M,1,2,\cdots,S$) to make inner products for each matching boundary, we obtain a matrix equation for the unknown coefficients:
 	\begin{align}
 			\label{Eq.mthmatMN}
 			\left(
 			\begin{array}{cccccccccccc}
 				\bm{M}^-_{\textrm{R}_0} & \bm{M}^+_{\textrm{T}_1} & \bm{M}^+_{\textrm{R}_1} & 0 & 0 &  & 0 & 0 & 0  \\
 				\bm{N}^-_{\textrm{R}_0} & \bm{N}^+_{\textrm{T}_1} & \bm{N}^+_{\textrm{R}_1} & 0 & 0 & \cdots & 0 & 0 & 0 \\
 				0 & \bm{M}^-_{\textrm{T}_1} & \bm{M}^-_{\textrm{R}_1} & \bm{M}^+_{\textrm{T}_2} & \bm{M}^+_{\textrm{R}_2} &  & 0 & 0 & 0  \\
 				0 & \bm{N}^-_{\textrm{T}_1} & \bm{N}^-_{\textrm{R}_1} & \bm{N}^+_{\textrm{T}_2} & \bm{N}^+_{\textrm{R}_2} &  & 0 & 0 & 0 \\
 			 	&             &      \vdots       &  &  & \ddots &&&&& \\
 				0 & 0 & 0 & 0 & 0 &  & \bm{M}^-_{\textrm{T}_N} & \bm{M}^-_{\textrm{R}_N} & \bm{M}^+_{\textrm{T}_{N+1}}\\
 				0 & 0 & 0 & 0 & 0 &  & \bm{N}^-_{\textrm{T}_N} & \bm{N}^-_{\textrm{R}_N} & \bm{N}^+_{\textrm{T}_{N+1}}
 			\end{array}
 			\right)
 			\left(
 			\begin{array}{c}
 				\bm{\alpha}_{\textrm{R}_0}\\
 				\bm{\alpha}_{\textrm{T}_1}\\
 				\bm{\alpha}_{\textrm{R}_1}\\
 				\bm{\alpha}_{\textrm{T}_2}\\
 				\bm{\alpha}_{\textrm{R}_2}\\
 				\vdots \\
 				\bm{\alpha}_{\textrm{T}_N}\\
 				\bm{\alpha}_{\textrm{R}_N}\\
 				\bm{\alpha}_{\textrm{T}_{N+1}}
 			\end{array}
 			\right) =
 			\left(
 			\begin{array}{c}
 				\bm{\beta}\\
 				\bm{\beta}_x\\
 				0\\
 				0\\
 				\vdots \\
 				0\\
 				0
 			\end{array}
 			\right).
 	\end{align}
 $\bm{M}^-_{\textrm{R}_0}$, $\bm{N}^-_{\textrm{R}_0}$, $\bm{M}^+_{\textrm{T}_{N+1}}$, and $\bm{N}^+_{\textrm{T}_{N+1}}$ are $M+S$ diagonal matrixes. If we use $Z_p$ to make an inner product for the potential of free-surface waves, the inner product will be orthogonal for different wave numbers {(see Meng and Lu \cite{2017-Meng-p567})}. The diagonal element is defined as follows
 	\begin{align}
 			&\bm{M}^-_{\textrm{R}_0}(p) = \mathscr{P}(p), & &\bm{N}^-_{\textrm{R}_0}(p) = \theta k_p \mathscr{P}(p),
 	\end{align}
 	\begin{align}
 			&\bm{M}^+_{\textrm{T}_{N+1}}(p) = - \mathscr{P}(p), & &\bm{N}^+_{\textrm{T}_{N+1}}(p) = \theta k_p \mathscr{P}(p),
 	\end{align}
 where
 	\begin{align}
 			&\mathscr{P}(p) = \left\langle Z_p,Z_p \right\rangle, & &\theta = \left\{
 			\begin{array}{cl}1, & (p=0_1,0_2,\cdots,0_M),\\ \textup{i}, & (p=1,2,\cdots,S).\end{array}\right.
 	\end{align}
 $\bm{M}^\pm_{\textrm{T}_n}$, $ \bm{M}^\pm_{\textrm{R}_n}$, $\bm{N}^\pm_{\textrm{T}_n}$, and $\bm{N}^\pm_{\textrm{R}_n}$ are $M+S$ by $M+S+2$ matrixes analogical to the matrixes in Eq.\ (\ref{Eq.mthMat}). We can also calculate every element by
 	\begin{align}
 			&\bm{M}^\pm_{\textrm{T}_n}(p,q) = \mp \textup{e}^{ \mp \textup{i} \delta \tilde{k}_{n,q} L_n} \mathscr{D}_n(p,q),
 	\end{align}
 	\begin{align}
 			&\bm{M}^\pm_{\textrm{R}_n}(p,q) = \mp \textup{e}^{ \pm \textup{i} \delta \tilde{k}_{n,q} L_n} \mathscr{D}_n(p,q),
 			\end{align}
 	\begin{align}
 			&\bm{N}^\pm_{\textrm{T}_n}(p,q) = \pm \delta \tilde{k}_{n,q} \textup{e}^{\mp \textup{i} \delta \tilde{k}_{n,q} L_n} \mathscr{D}_n(p,q),
 	\end{align}
 	\begin{align}
 			 &\bm{N}^\pm_{\textrm{R}_n}(p,q) = \mp \delta \tilde{k}_{n,q} \textup{e}^{\pm \textup{i} \delta \tilde{k}_{n,q} L_n} \mathscr{D}_n(p,q),
 	\end{align}
 where
 	\begin{align}
 			\delta =
 			\left\{
 			\begin{array}{cl}
 					1, & (q=0_1,0_2,\cdots,0_M),\\
 					\textup{i}, & (q=\textup{I},\textup{II},1,2,\cdots,S),
 			\end{array}
 			\right.
 	\end{align}	
 $\bm{\beta}$ and $\bm{\beta}_x$ are $(M+S)$-dimensional column vectors related to the incident waves:
 	\begin{align}
 			&\bm{\beta} = \left[ -I_{0,0_1}\mathscr{P}(0_1), \cdots, -I_{0,0_M}\mathscr{P}(0_M),0,0,\cdots\right]^\intercal,
 	\end{align}
 	\begin{align}
 			&\bm{\beta}_x = \left[ k_{0_1} I_{0,0_1} \mathscr{P}(0_1), \cdots, k_{0_M} I_{0,0_M} \mathscr{P}(0_M),0,0,\cdots\right]^\intercal.
 	\end{align}
 The undetermined coefficient vectors are
 	\begin{align}
 			&\bm{\alpha}_{\textrm{R}_0} = \left[ R_{0,0_1},\cdots,R_{0,0_M},R_{0,1},\cdots,R_{0,S}\right]^\intercal,
 	\end{align}
 	\begin{align}
 			&\bm{\alpha}_{\textrm{T}_n} = \left[ T_{n,0_1},\cdots,T_{n,0_M},T_{n,\textup{I}},T_{n,\textup{II}},T_{n,1},\cdots,T_{n,S}\right]^\intercal,
 	\end{align}
 	\begin{align}
 			&\bm{\alpha}_{\textrm{R}_n} = \left[ R_{n,0_1},\cdots,R_{n,0_M},R_{n,\textup{I}},R_{n,\textup{II}},R_{n,1},\cdots,R_{n,S}\right]^\intercal,
 	\end{align}
 	\begin{align}
 			&\bm{\alpha}_{\textrm{T}_{N+1}} = \left[ T_{N+1,0_1},\cdots,T_{N+1,0_M},T_{N+1,1},\cdots,T_{N+1,S}\right]^\intercal.
 	\end{align}

 Additional connection conditions are yet {required} to complete a closed equation system for the calculation. For the generalized situation, arbitrary forms of the connection conditions are practicable to the group of elastic plates. {We choose the elastic torsion joint as a typical instance of connecting types to elaborate the process.} Assuming the conditions at $x=a_0$ and $x=a_N$ are free edge conditions and the discontinuity at $x=a_n$ ($n=1,2,\cdots,N-1$) is a rotational spring with torsional rigidity $J_n$, we write the conditions as follows:
	\begin{align}
			\label{Eq.edgCon1}
			\frac{\textup{d}^2 \zeta}{\textup{d} x^2} = \frac{\textup{d}^3 \zeta}{\textup{d} x^3} = 0, && (x=a_0,a_N),
	\end{align}
	\begin{align}
			\label{Eq.hinCon1}
			\zeta \Big|_{x=a_n^-} = \zeta \Big|_{x=a_n^+}, && (n=1,2,\cdots,N-1),
	\end{align}
	\begin{align}
			\label{Eq.hinCon2}
			D_n \frac{\textup{d}^2 \zeta}{\textup{d} x^2} \Big|_{x=a_n^-} = D_{n+1} \frac{\textup{d}^2 \zeta}{\textup{d} x^2} \Big|_{x=a_n^+} , && (n=1,2,\cdots,N-1),
	\end{align}
	\begin{align}
			\label{Eq.hinCon3}
			D_n \frac{\textup{d}^2 \zeta}{\textup{d} x^2} \Big|_{x=a_n^-} = J_n \left[ \frac{\textup{d} \zeta}{\textup{d} x} \Big|_{x=a_n^+}- \frac{\textup{d} \zeta}{\textup{d} x} \Big|_{x=a_n^-} \right], && (n=1,2,\cdots,N-1),
	\end{align}
	\begin{align}
			\label{Eq.hinCon4}
			D_n \frac{\textup{d}^3 \zeta}{\textup{d} x^3} \Big|_{x=a_n^-} = D_{n+1} \frac{\textup{d}^3 \zeta}{\textup{d} x^3}\Big|_{x=a_n^+}, && (n=1,2,\cdots,N-1).
	\end{align}
 By applying Eqs.\ (\ref{Eq.potExpTnExt}) and (\ref{Eq.potExpRnExt}) to Eqs.\ (\ref{Eq.edgCon1})--(\ref{Eq.hinCon4}), we have
 	\begin{align}
 			\label{Eq.edgmatMN}
 			\left(
 			\begin{array}{ccccccccccc}
 					\bm{E}^+_{\textrm{T}_1} & \bm{E}^+_{\textrm{R}_1} & 0 & 0 & 0 & 0 & & 0 & 0 & 0 & 0\\
 					\bm{E}^-_{\textrm{T}_1} & \bm{E}^-_{\textrm{R}_1} & \bm{E}^+_{\textrm{T}_2} & \bm{E}^+_{\textrm{R}_2} & 0 & 0 &\cdots  & 0 & 0 & 0 & 0\\
 					0 & 0 & \bm{E}^-_{\textrm{T}_2} & \bm{E}^-_{\textrm{R}_2} & \bm{E}^+_{\textrm{T}_3} & \bm{E}^+_{\textrm{R}_3} & & 0 & 0 & 0 & 0 \\
 					&  & \vdots & &  &  & \ddots &  &  &  \\
 					0 & 0 & 0 & 0 & 0 & 0 &  & \bm{E}^-_{\textrm{T}_{N-1}} & \bm{E}^-_{\textrm{R}_{N-1}} & \bm{E}^+_{\textrm{T}_N} & \bm{E}^+_{\textrm{R}_N}\\
 					0 & 0 & 0 & 0 & 0 & 0 &  & 0 & 0 & \bm{E}^-_{\textrm{T}_N} & \bm{E}^-_{\textrm{R}_N}
 			\end{array}
 			\right)
 			\left(
 			\begin{array}{c}
 			\bm{\alpha}_{\textrm{T}_1}\\
 			\bm{\alpha}_{\textrm{R}_1}\\
 			\bm{\alpha}_{\textrm{T}_2}\\
 			\bm{\alpha}_{\textrm{R}_2}\\
 			\vdots\\
 			\bm{\alpha}_{\textrm{T}_N}\\
 			\bm{\alpha}_{\textrm{R}_N}
 			\end{array}
 			\right) = \bm{0}.
 	\end{align}	
 $\bm{E}^+_{\textrm{T}_1}$, $\bm{E}^+_{\textrm{R}_1}$, $\bm{E}^-_{\textrm{T}_N}$, and $\bm{E}^-_{\textrm{R}_N}$ are $2$ by $M+S+2$ matrixes related to the sequence $q$, where every column has the form of
 	\begin{align}
 			&\bm{E}^+_{\textrm{T}_1}(q) =
 					\left(
 					\begin{array}{c}
 					\delta^2 \tilde{k}_{1,q}^2 W^-_{1,q}  \\
 					\delta^3 \tilde{k}_{1,q}^3 W^-_{1,q}
 					\end{array}
	 				\right),
	 		& &\bm{E}^+_{\textrm{R}_1}(q) =
 					\left(
 					\begin{array}{c}
 					\delta^2 \tilde{k}_{1,q}^2 W^+_{1,q}  \\
 					-\delta^3 \tilde{k}_{1,q}^3 W^+_{1,q}
 					\end{array}
	 				\right),
	\end{align}
	\begin{align}
	 		&\bm{E}^-_{\textrm{T}_N}(q) =
 					\left(
 					\begin{array}{c}
 					\delta^2 \tilde{k}_{N,q}^2 W^+_{N,q}  \\
 					\delta^3 \tilde{k}_{N,q}^3 W^+_{N,q}
 					\end{array}
	 				\right),
	 		& &\bm{E}^-_{\textrm{R}_N}(q) =
 					\left(
 					\begin{array}{c}
 					\delta^2 \tilde{k}_{N,q}^2 W^-_{N,q}  \\
 					-\delta^3 \tilde{k}_{N,q}^3 W^-_{N,q}
 					\end{array}
	 				\right).
	\end{align}
 $\bm{E}^-_{\textrm{T}_n}$, $\bm{E}^-_{\textrm{R}_n}$ ($n=1,\cdots,N-1$) and $\bm{E}^+_{\textrm{T}_n}$, $\bm{E}^+_{\textrm{R}_n}$ ($n=2,\cdots,N$) are $4$ by $M+S+2$ matrixes and the expression for every column reads
 	\begin{align}
 			&\bm{E}^-_{\textrm{T}_n}(q) =
 					\left(
 					\begin{array}{c}
 					W^+_{n,q} \\
 					\delta^2 \tilde{k}^2_{n,q} D_n W^+_{n,q} \\
 					(\delta^2 \tilde{k}^2_{n,q} D_n - \textup{i} \delta \tilde{k}_{n,q} J_n) W^+_{n,q} \\
 					\delta^3 \tilde{k}^3_{n,q} D_n W^+_{n,q}
 					\end{array}
 					\right),
 			& &\bm{E}^-_{\textrm{R}_n}(q)  =
 					\left(
 					\begin{array}{c}
 					W^-_{n,q} \\
 					\delta^2 \tilde{k}^2_{n,q} D_n W^-_{n,q}\\
 					(\delta^2 \tilde{k}^2_{n,q} D_n + \textup{i} \delta \tilde{k}_{n,q} J_n) W^-_{n,q}\\
 					-\delta^3 \tilde{k}^3_{n,q} D_n W^-_{n,q}
 					\end{array}
 					\right),
 	\end{align}
 	\begin{align}
 			&\bm{E}^+_{\textrm{T}_n}(q)  =
 					\left(
 					\begin{array}{c}
 					- W^-_{n,q}\\
 					- \delta^2 \tilde{k}^2_{n,q} D_n W^-_{n,q}\\
 					\textup{i} \delta \tilde{k}_{n,q} J_{n-1} W^-_{n,q}\\
 					-\delta^3 \tilde{k}^3_{n,q} D_n W^-_{n,q}
 					\end{array}
 					\right),
 			& &\bm{E}^+_{\textrm{R}_n}(q)  =
 					\left(
 					\begin{array}{c}
 					- W^+_{n,q}\\
 					- \delta^2 \tilde{k}^2_{n,q} D_n W^+_{n,q}\\
 					- \textup{i} \delta \tilde{k}_{n,q} J_{n-1} W^+_{n,q}\\
 					\delta^3 \tilde{k}^3_{n,q} D_n W^+_{n,q}
 					\end{array}
 					\right),
 	\end{align}
 where
 	\begin{align}
 			W^\pm_{n,q} = \textup{e}^{\pm \textup{i} \delta \tilde{k}_{n,q} L_n} \dfrac{\partial \tilde{Z}_{n,q}}{\partial z}\Big|_{z=0}.
 	\end{align}

 Equations (\ref{Eq.mthmatMN}) and (\ref{Eq.edgmatMN}) consist of $2(M+S)(N+1) + 4N$ simultaneous equations for $2(M+S)(N+1) + 4N$ unknown coefficients, thus the generalized problem can be solved afterwards.

 The formulations for the method above are also appropriate for many other situations after simple and minor modifications. For example, a three-dimensional problem of waves incoming in oblique directions (e.g.\ Williams and Squire \cite{2006-Williams-p113}) can be solved after a decomposition for wave numbers by Snell's law. The simultaneous equations for an infinite ice sheet with undulant topography (e.g.\ Porter and Porter \cite{2004-Porter-p145}, Williams and Squire \cite{2004-Williams-p3469}) may be obtained as well if one uses a stepwise segmentation to approximate the continuously changing profile of the ice sheet. The inner product technique is very eligible to deal with the matching relations for a problem of linearity, whether in a Cartesian coordinate system or in a polar one. {{More possibilities should be explored for more sophisticated cases.}}

\section{\label{Sec.4.Calcu}Calculations and discussions}

\subsection{\label{subSec.4.1.conValida}Convergence Validation}

 Let $\rho_1$, $H_M$ and $\sqrt{H_M/g}$ be respectively the characteristic quantities of density, length and time to transfer the problem to a nondimensionalized system. Given that the oscillations of the group of elastic plates and the connections are {conservative}, i.e.\ the mechanical energy have no dissipation, we can deduce an energy conservation relation by use of Green's theorem, which takes the form of
 \begin{align}
		\label{Eq.eneCon}
		\Delta = \Delta_1 + \Delta_2 + \cdots + \Delta_M \equiv 0,
 \end{align}
 where
 \begin{align}
 		\Delta_m = k_{0_m} \mathscr{P}(0_m) \left( \lVert I_{0,0_m} \rVert^2 - \lVert R_{0,0_m} \rVert^2 - \lVert T_{N+1,0_m} \rVert^2 \right).
 \end{align}

 $\Delta_m$ can be regarded as the residual energy of the incident wave of mode $k_{0_m}$ after being reflected and transmitted to the far field. A positive $\Delta_m$ indicates that {the remainder part of the energy} of $k_{0_m}$ will be transferred to the other modes after scattering, while for a negative one, the wave motion of $k_{0_m}$ will receive energy form the others. The evanescent modes and the group of the plates {serve as the agency through the whole} course of energy exchange. For a reversible structure floating on an $M$-layer fluid, Eq.\ (\ref{Eq.eneCon}) reveals the law of energy transfer between the motions of each wave mode, which is also an effective {verification for} the convergence of numerical calculations.

 As {a verification}, we compute two elastic plates with a torsion joint floating on a 2-layer fluid, i.e.\ let $M=N=2$. The constants used are $D_1 = D_2 = 0.05$, $M_1 = M_2 = 0.0001$, $ L_1 = L_2 = 6$, $J_1 = 0.05$, $h_1=0.2$, $h_2=0.8$, $\xi_1 = 0.01$, $\xi_2 = 0.0001$, and $\gamma_1 = 0.9$. The conservation expression $\Delta$ and the relative error $\varepsilon$ related to $\Delta_1$ and $\Delta_2$ are listed in Tab.\ \ref{Tab.convg} as follows. In Tab.\ \ref{Tab.convg}, $\varepsilon$ is define by
 \begin{align}
 		\varepsilon = \dfrac{\lvert\lvert\Delta_1\rvert-\lvert\Delta_2\rvert\rvert}{\textup{min}\left(\lvert\Delta_1\rvert,\lvert\Delta_2\rvert\right)} \times 100\%.
 \end{align}
 \begin{table}[H]
 	  \caption{{Computation  convergence for different} truncating numbers $S$, where $\omega=0.8,1.6,2.4$}	
	  \label{Tab.convg}
	  \centering
	  \begin{tabular*}{\textwidth}{p{4cm}p{4cm}p{5.5cm}p{4cm}}
			  \toprule
			  $\omega$    & $S$      & $\Delta$    & $\varepsilon$\\
			  \midrule
			   $0.8$      & $5 $      & $ 7.182350 \times 10^{-14} $  & $ 0.148\% $  \\
			              & $15$      & $ 5.782463 \times 10^{-15} $  & $ 0.012\% $  \\
			              & $25$      & $ 1.514144 \times 10^{-15} $  & $ 0.003\% $  \\

   			   $1.6$      & $5 $      & $ 3.043093 \times 10^{-15} $  & $ 4.570\% $  \\
			              & $15$      & $ 4.080502 \times 10^{-16} $  & $ 0.597\% $  \\
			              & $25$      & $ 1.205577 \times 10^{-16} $  & $ 0.176\% $  \\

			   $2.4$      & $5 $      & $ 3.872847 \times 10^{-20} $  & $ 15.990\% $  \\
			              & $15$      & $ 8.860126 \times 10^{-21} $  & $ 3.444\% $  \\
			              & $25$      & $ 3.463698 \times 10^{-21} $  & $ 1.327\% $  \\
			  \bottomrule
	  \end{tabular*}
 \end{table}

 It is distinct that the method will provide a great converging rate. For a lower frequency, only 5 truncating terms are required to assure the convergence, while for a higher frequency, 25 terms are {already sufficient} to yield a convincible result. {The {high computational efficiency can also be checked by} considering the continuities of pressure and velocity along each matching boundary, and a coincident rate of convergence must be exhibited.}

 \subsection{\label{subSec.4.2.compKohout}Result comparison with previous analytical method}

\begin{figure}[H]
		\centering
		\makebox[0.9\textwidth][l]{($a$)}
		\includegraphics[width=\textwidth]{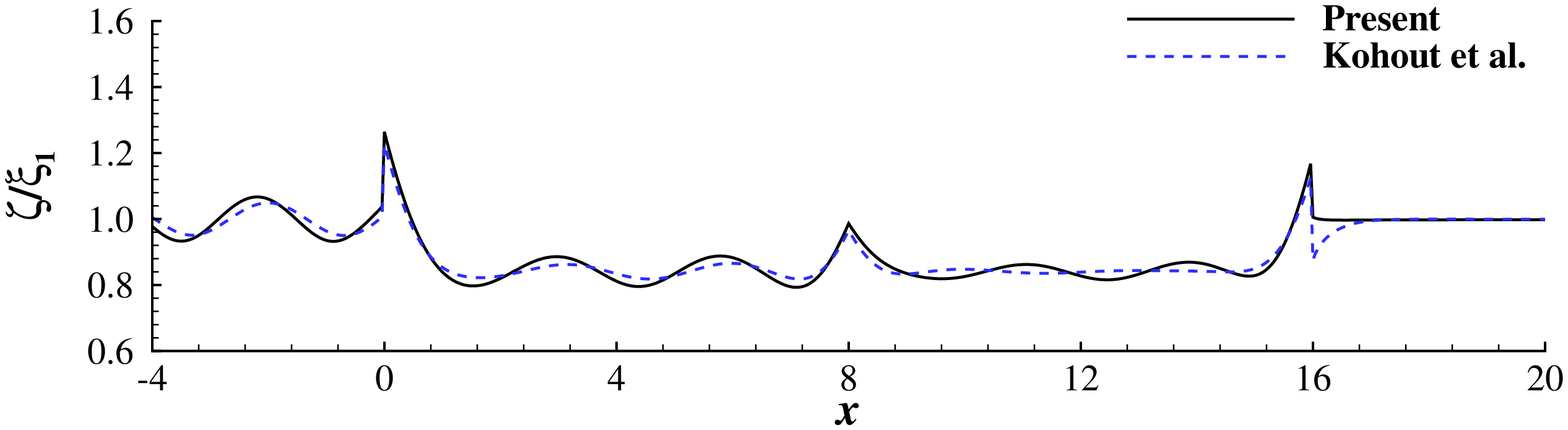}\\
		\makebox[0.9\textwidth][l]{($b$)}
		\includegraphics[width=\textwidth]{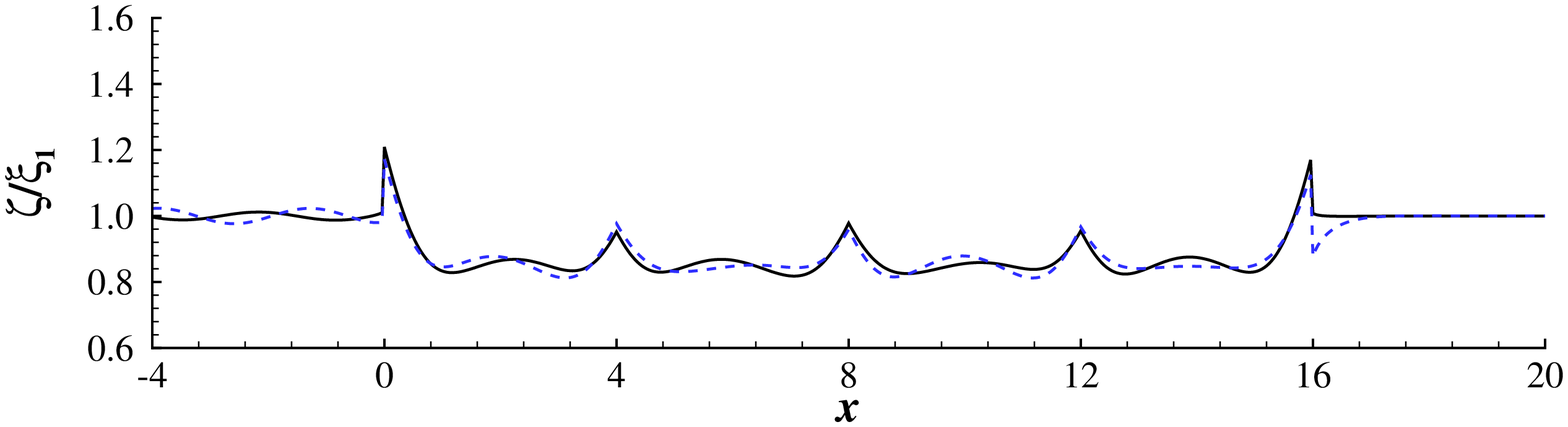}
		\caption{Displacement amplitude on surface calculated by present method and Kohout et al. \cite{2007-Kohout-p649} with ($a$) $N=2$, ($b$) $N=4$}
		\label{Fig.dis_methCom}
 \end{figure}

\begin{figure}[H]
		\centering
		\makebox[0.9\textwidth][l]{($a$)}
		\includegraphics[width=\textwidth]{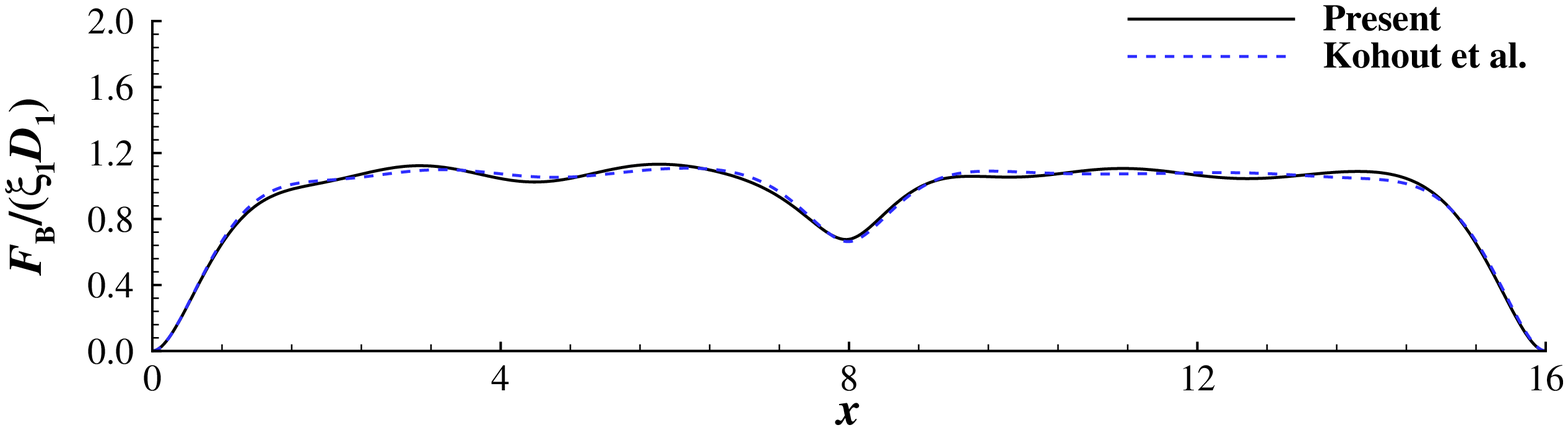}\\
		\makebox[0.9\textwidth][l]{($b$)}
		\includegraphics[width=\textwidth]{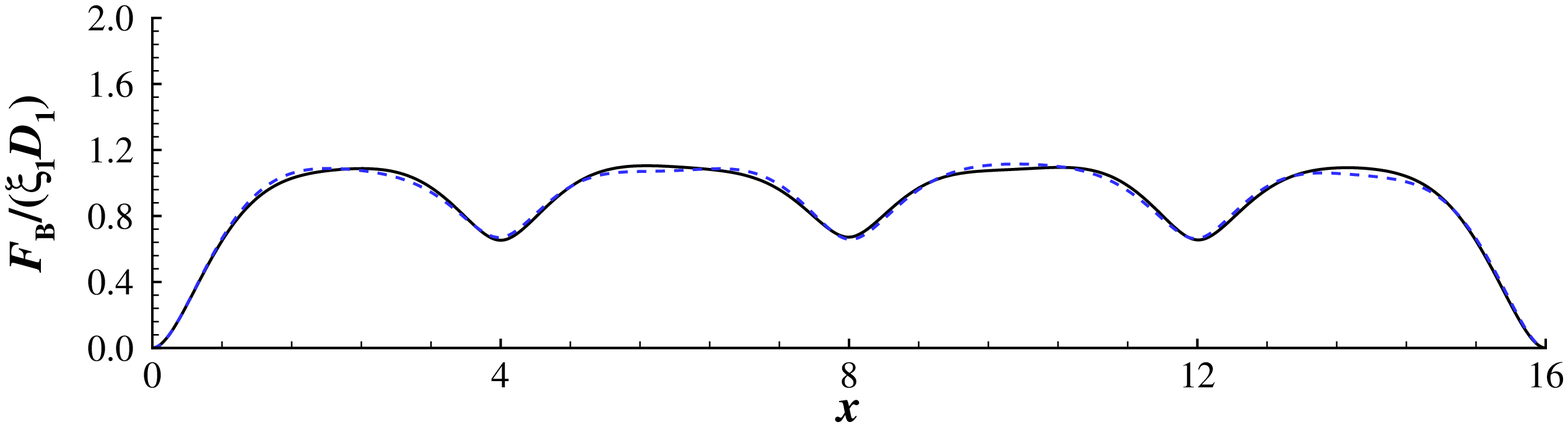}
		\caption{Amplitude of the bending moment calculated by present method and Kohout et al. \cite{2007-Kohout-p649} with ($a$) $N=2$, ($b$) $N=4$}
		\label{Fig.bdm_methCom}
 \end{figure}

\begin{figure}[H]
		\centering
		\makebox[0.9\textwidth][l]{($a$)}
		\includegraphics[width=\textwidth]{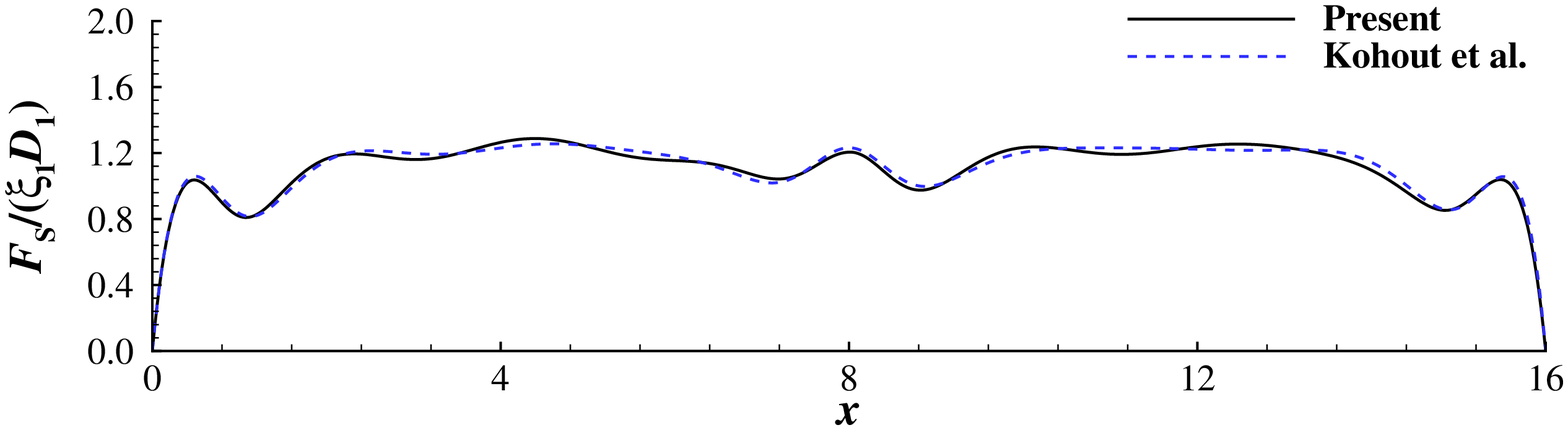}\\
		\makebox[0.9\textwidth][l]{($b$)}
		\includegraphics[width=\textwidth]{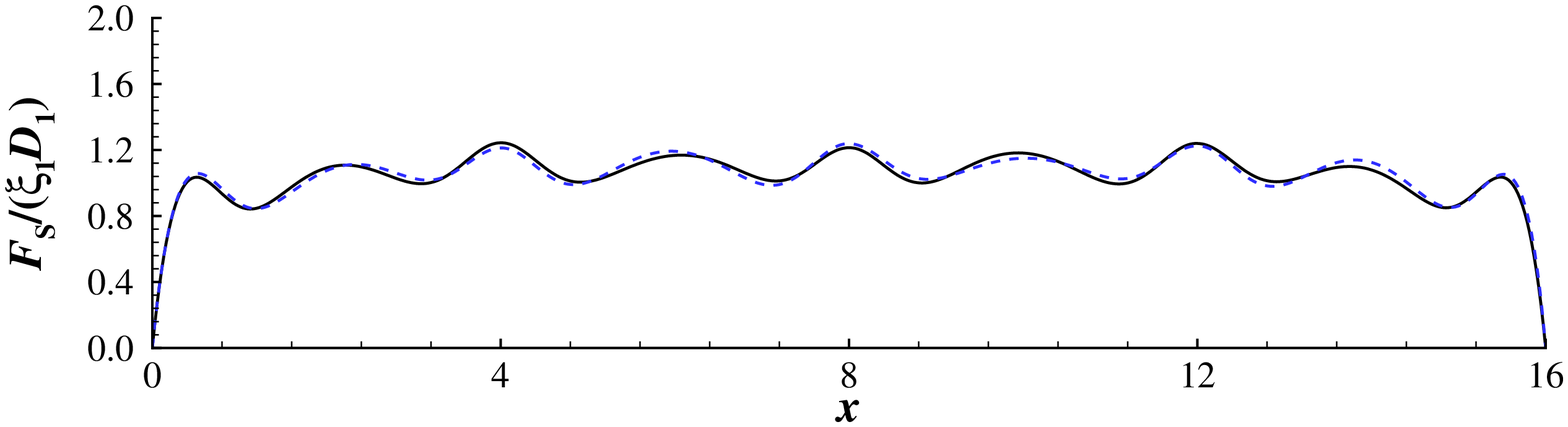}
		\caption{Amplitude of the shear force calculated by present method and Kohout et al. \cite{2007-Kohout-p649} with ($a$) $N=2$, ($b$) $N=4$}
		\label{Fig.shf_methCom}
 \end{figure}

  We carry out a comparison between the present method and the method of Kohout et al.\ \cite{2007-Kohout-p649} to validate the correctness of our calculation. Although the inner product technique described in Kohout et al.\ \cite{2007-Kohout-p649} was employed to study the three-dimensional problem that obliquely incident waves propagate across infinite inhomogeneous ice floes with multiple cracks floating on a uniform fluid, it is also eligible to study an analogous two-dimensional problem. {A convincible argument} is that the inner product system deals with the matching relations of pressure and of velocity with respect to the vertical direction, and the vertical eigenfunctions {will not be affected by} different incident angles. Figures \ref{Fig.dis_methCom}--\ref{Fig.shf_methCom} show the calculation comparisons for the surface displacement, the bending moment and the shear force of the group of plates by these two methods. The fluid is of uniform density. For the incident wave, the frequency and amplitude are assigned as $\omega = 1$ and $\xi_1 = 0.01 $; {the constants for the physical properties of every single plate and the torsion joint} are $D_1 = \cdots = D_4 = 0.05$, $M_1 = \cdots = M_4 = 0.0001$, and $J_1 = J_2 = J_3 = 0.05$. To emphasize the virtue of the methods in manipulating matching boundaries between the multiple domains covered by elastic plates, we set up two structure models with $N=2$, $L_1 = L_2 = 4$ and $N=4$, $L_1 = \cdots = L_4 = 2$. It is conspicuous that the present method {exhibits globally identical results} with the previous method of Kohout et al.\ \cite{2007-Kohout-p649}.

 \subsection{\label{subSec.4.3.mulPlate}Group of elastic plates subjected to incident waves}

  {The hydrodynamic responses of a floating structure composed by different combinations of multiple elastic plates are considered hereinafter.} Three sorts of combinations of elastic plates and torsion springs are evaluated for incident waves in a 2-layer fluid, where $M=2$, $\xi_1 = 0.01$, $\xi_2 = 0.0001$, $h_1 = 0.2$, $h_2 = 0.8$, and $\gamma_1 = 0.9$. The displacements, the bending moments and the shear forces of the structure are considered for different incident frequencies. The total length of the structure is limited to a constant, and the number of the plates and the distribution of the connections are configured as follows: for $N = 1$, $L_1 = 8$; for $N = 2$, $L_1 = L_2 = 4$; for $N = 4$, $L_1 = \cdots = L_4 = 2$. The physical properties for every single plate and connection are consistent with those in Sec.\ 4.2.
 
 % \ref{subSec.4.2.compKohout}.

\begin{figure}[H]
		\centering
		\makebox[0.9\textwidth][l]{($a$)}
		\includegraphics[width=\textwidth]{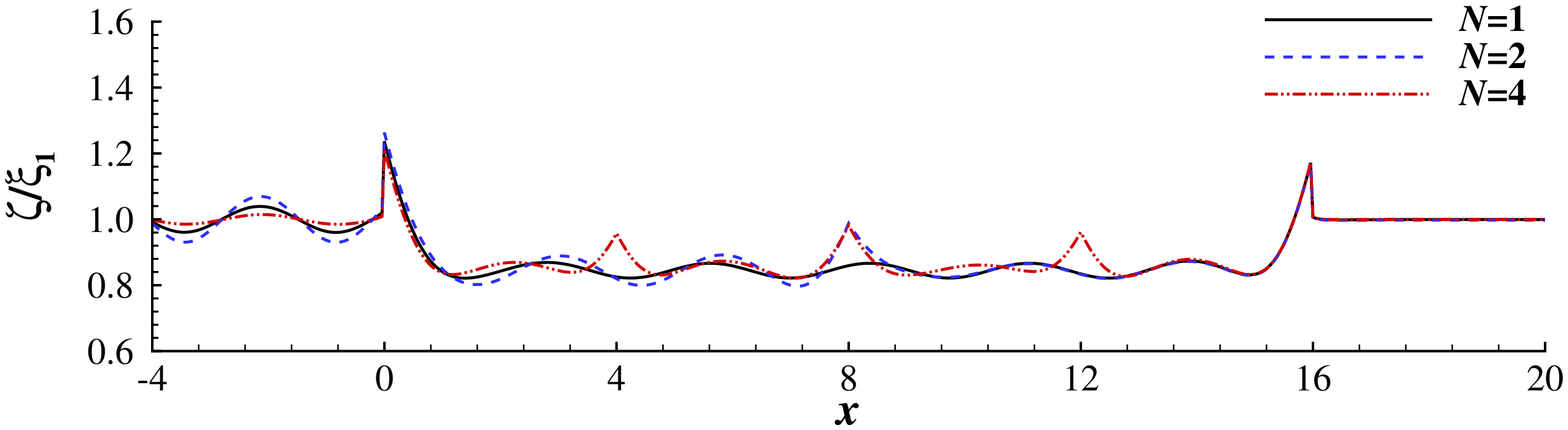}\\
		\makebox[0.9\textwidth][l]{($b$)}
		\includegraphics[width=\textwidth]{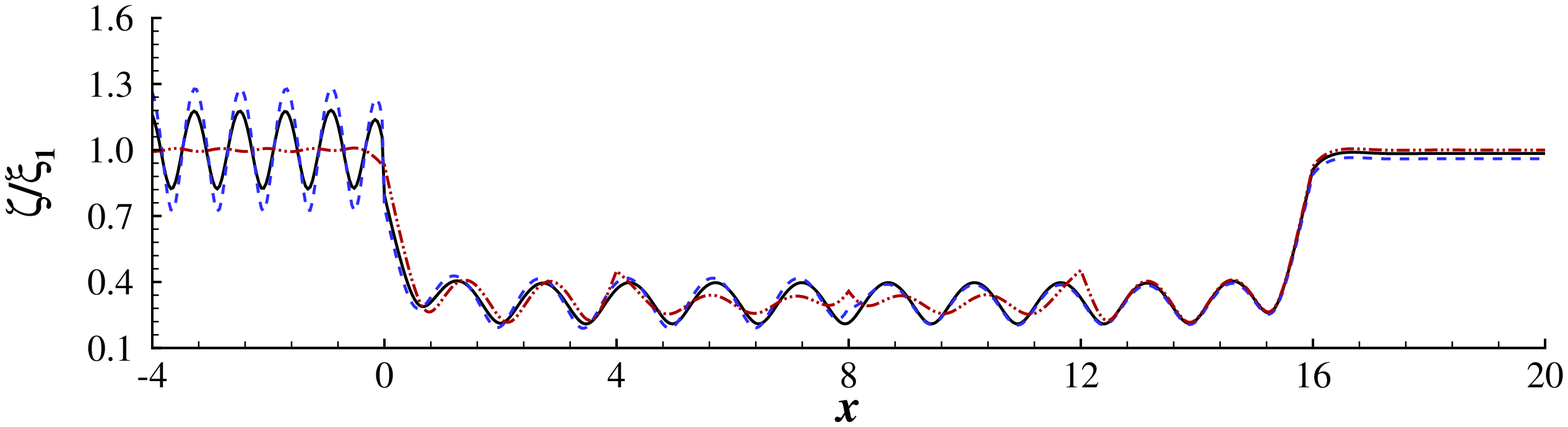}
		\caption{Displacement amplitude on surface by different combinations of elastic plates for ($a$) $\omega = 1$, ($b$) $\omega = 2$}
		\label{Fig.dis_omeg_N124}
 \end{figure}

\begin{figure}[H]
		\centering
		\makebox[0.9\textwidth][l]{($a$)}
		\includegraphics[width=\textwidth]{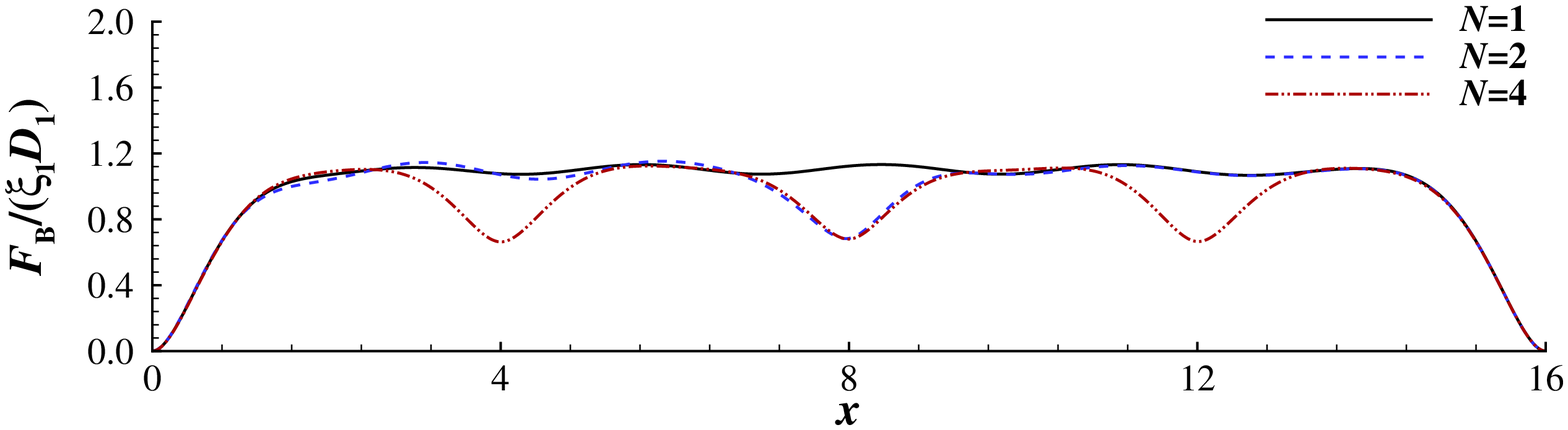}\\
		\makebox[0.9\textwidth][l]{($b$)}
		\includegraphics[width=\textwidth]{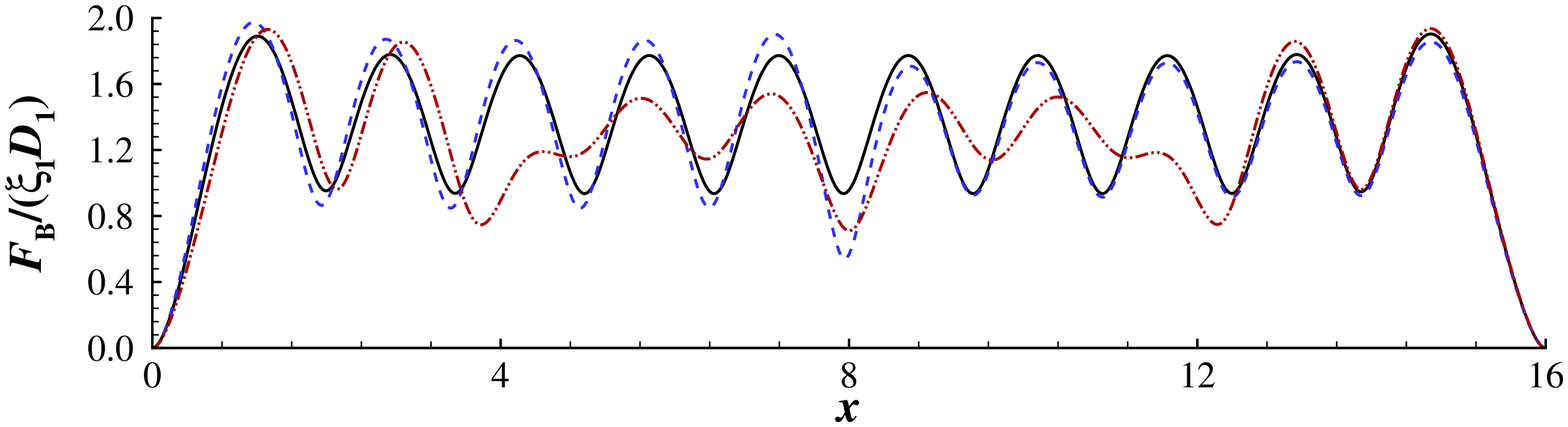}
		\caption{Amplitude of the bending moment by different combinations of elastic plates for ($a$) $\omega = 1$, ($b$) $\omega = 2$}
		\label{Fig.bdm_omeg_N124}
 \end{figure}

\begin{figure}[H]
		\centering
		\makebox[0.9\textwidth][l]{($a$)}
		\includegraphics[width=\textwidth]{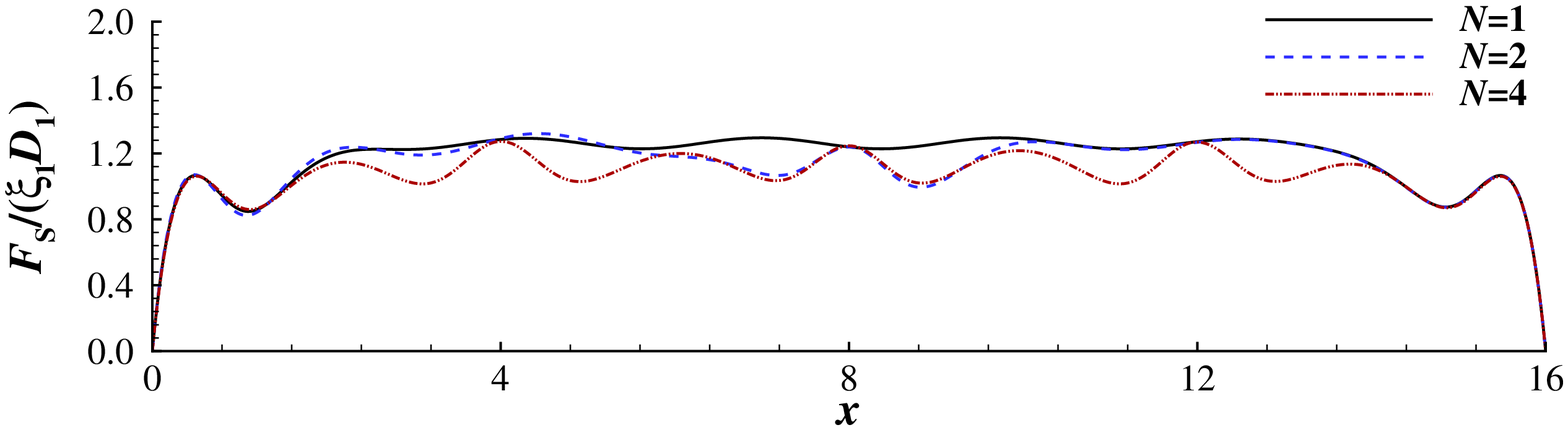}\\
		\makebox[0.9\textwidth][l]{($b$)}
		\includegraphics[width=\textwidth]{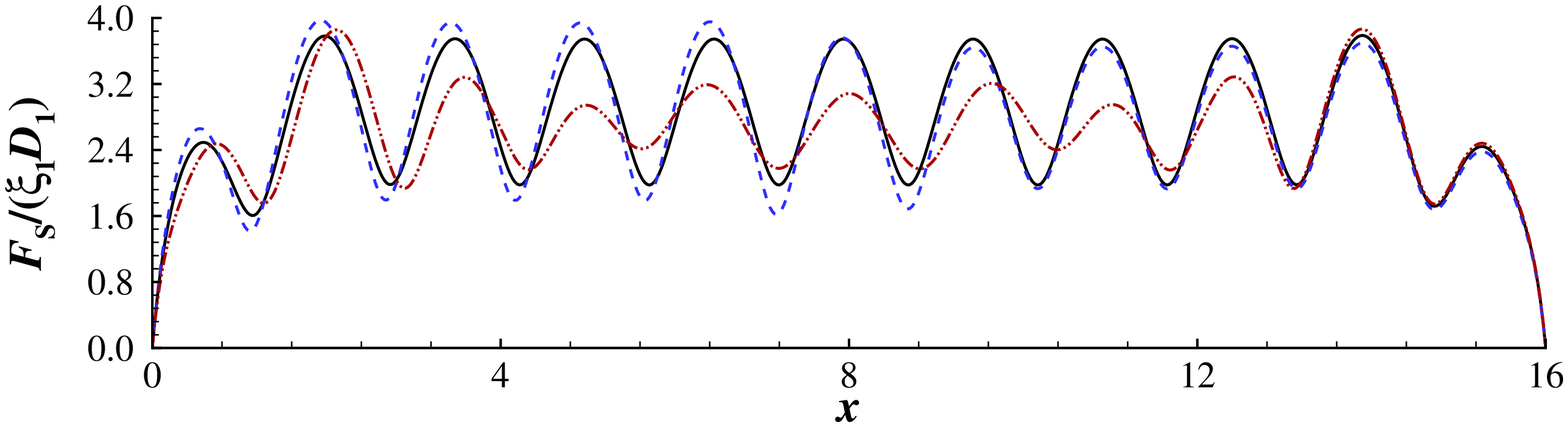}
		\caption{Amplitude of the shear force by different combinations of elastic plates for ($a$) $\omega = 1$, ($b$) $\omega = 2$}
		\label{Fig.shf_omeg_N124}
 \end{figure}

 Considering the situation of reflection and transmission in Fig.\ \ref{Fig.dis_omeg_N124}, it is noted that, for a lower frequency, namely $\omega=1$, the maximum of displacement amplitudes in the region $x<0$ is on the curve of the 2-plate model, while for $\omega=2$, it is on the 4-plate model. {The reflection capacity of the multi-module structure is affected by the combinations of elastic plates for different frequency stages.} Along $0<x<16$ in Figs.\ \ref{Fig.dis_omeg_N124}--\ref{Fig.shf_omeg_N124}, the amplitude profiles of the deflections, bending moments and shear forces also show disagreements for various structures. For $\omega=1$, the deflection amplitudes by the three combinations are almost in the same level except for the neighborhoods nearby the connections, where the elastic torsion springs oscillate intensely stronger. For the bending moments, the amplitude values nearby the connections exhibit a concave profile for the composite structures, i.e.\ 2-plate and 4-plate models; the shear forces of the composite structures are lower than the single one. For $\omega=2$, the curves of the 4-plate model for deflection, bending moment and shear force have the smallest maximum values in the region $4<x<12$.

 \subsection{\label{subSec.4.4.varDen}Group of elastic plates floating on stratified fluids}

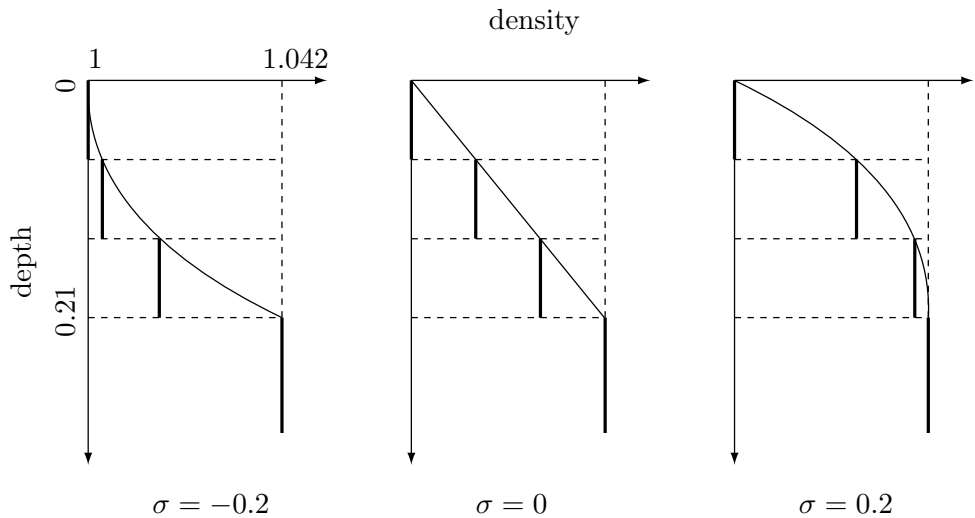
\begin{figure}[H]
    	\centering
    	\setlength{\unitlength}{0.85cm}
    	\begin{picture}(14,7)
    			\put(6.2,6.8){density}
    			\put(-1.2,2.6){\rotatebox{90}{depth}}
    			\put(0,6.2){1}
    			\put(2.7,6.2){1.042}
    			\put(-0.5,5.8){\rotatebox{90}{0}}
    			\put(-0.5,2){\rotatebox{90}{0.21}}
    			% one
				\linethickness{0.05\unitlength}
				\put(3,2.3){\line(0,-1){1.8}}
				\put(0,6){\line(0,-1){1.233}}
				\put(0.22,4.766){\line(0,-1){1.233}}
				\put(1.1,3.533){\line(0,-1){1.233}}
				\linethickness{0.02\unitlength}
				\put(0,6){\vector(1,0){3.7}}
				\put(0,6){\vector(0,-1){6}}
				\multiput(0,3.533)(0.2,0){15}{\line(1,0){0.1}}
				\multiput(0,4.766)(0.2,0){15}{\line(1,0){0.1}}
				\multiput(0,2.3)(0.2,0){15}{\line(1,0){0.1}}
				\multiput(3,6)(0,-0.2){19}{\line(0,-1){0.1}}
				\qbezier(0,6)(-0.1,3.8)(3,2.3)
				\put(1,-0.75){$\sigma=-0.2$}
				% two
				\linethickness{0.05\unitlength}
				\put(8,2.3){\line(0,-1){1.8}}
				\put(5,6){\line(0,-1){1.233}}
				\put(6,4.766){\line(0,-1){1.233}}
				\put(7,3.533){\line(0,-1){1.233}}
				\linethickness{0.02\unitlength}
				\put(5,6){\vector(1,0){3.7}}
				\put(5,6){\vector(0,-1){6}}
				\multiput(5,3.53)(0.2,0){15}{\line(1,0){0.1}}
				\multiput(5,4.766)(0.2,0){15}{\line(1,0){0.1}}
				\multiput(5,2.3)(0.2,0){15}{\line(1,0){0.1}}
				\put(5,6){\line(0.121,-0.15){3}}
				\multiput(8,6)(0,-0.2){19}{\line(0,-1){0.1}}
				\put(6,-0.75){$\sigma=0$}
				% three
				\linethickness{0.05\unitlength}
				\put(13,2.3){\line(0,-1){1.8}}
				\put(10,6){\line(0,-1){1.233}}
				\put(11.89,4.766){\line(0,-1){1.233}}
				\put(12.79,3.533){\line(0,-1){1.233}}
				\linethickness{0.02\unitlength}
				\put(10,6){\vector(1,0){3.7}}
				\put(10,6){\vector(0,-1){6}}
				\multiput(10,3.53)(0.2,0){15}{\line(1,0){0.1}}
				\multiput(10,4.766)(0.2,0){15}{\line(1,0){0.1}}
				\multiput(10,2.3)(0.2,0){15}{\line(1,0){0.1}}
				\multiput(13,6)(0,-0.2){19}{\line(0,-1){0.1}}
				\qbezier(10,6)(13.1,4.5)(13,2.3)
				\put(11,-0.75){$\sigma=0.2$}
		\end{picture}
		\vspace{3mm}
	    \caption{Discrete density distribution for a 4-layer stratification depending on a parabolic curve}
		\label{Fig.schDiaForDensity}
	\end{figure}

 The water density of real ocean circumstance is inevitably stratified along the depth due to the unbalanced distribution of temperature and salinity. {It is especially obvious in the upper area of seawater, where usually exists a density pycnocline.} In order to study the impact of stratification on the mechanical behavior of a group of floating elastic plates, we try to use a parabolic curve to {represent the density versus depth in the upper area and use a constant to denote a uniform layer below. By taking several points on the curve as approximation, the discrete hierarchical system for the fluid is established, as shown in Fig.\ \ref{Fig.schDiaForDensity}. A 4-layer and an 8-layer fluid are configured for the calculation.}
 The density $\rho_m$ versus the depth $H_{m-1}$ ($m=1,\cdots,M$) follows the parabolic function as follows
 \begin{align}
 		\label{Eq.denVsDep}
 		\rho_m = - 4.76 \sigma H_{m-1}^2 + \left(\sigma + 0.2 \right) H_{m-1} + 1, && \left(-0.2 \le \sigma \le 0.2 \right),
 \end{align}
 where the parameter $\sigma$ is employed to simulate the profile of the curve. Once the stratification is {determined by $\sigma$, the density of each layer will be obtained.} In the computing process, the 4-layer fluid is separated by the constants $h_1 = h_2 = h_3 = 0.07$ and $h_4 = 0.79$, while the 8-layer fluid is by $h_1 = \cdots = h_7 = 0.03$ and $h_8 = 0.79$. The studied structure is a group of four elastic plates connected by torsion springs, with constants configured as follows $N=4$, $D_1= \cdots = D_4 = 0.05$, $M_1 = \cdots = M_4 = 0.0001$, $L_1 = \cdots = L_4 = 2$, and $J_1 = J_2 = J_3 = 0.05$. The frequency of incident waves is $\omega=0.2$. In order to avoid the interference of internal waves, we assume the incident waves only propagate in the surface traveling mode. Let $\xi_1 = 0.01$ and $\xi_n = 0$ $(n =2,3,\cdots,N)$.

\begin{figure}[H]
		\centering
		\makebox[0.9\textwidth][l]{($a$)}
		\includegraphics[width=\textwidth]{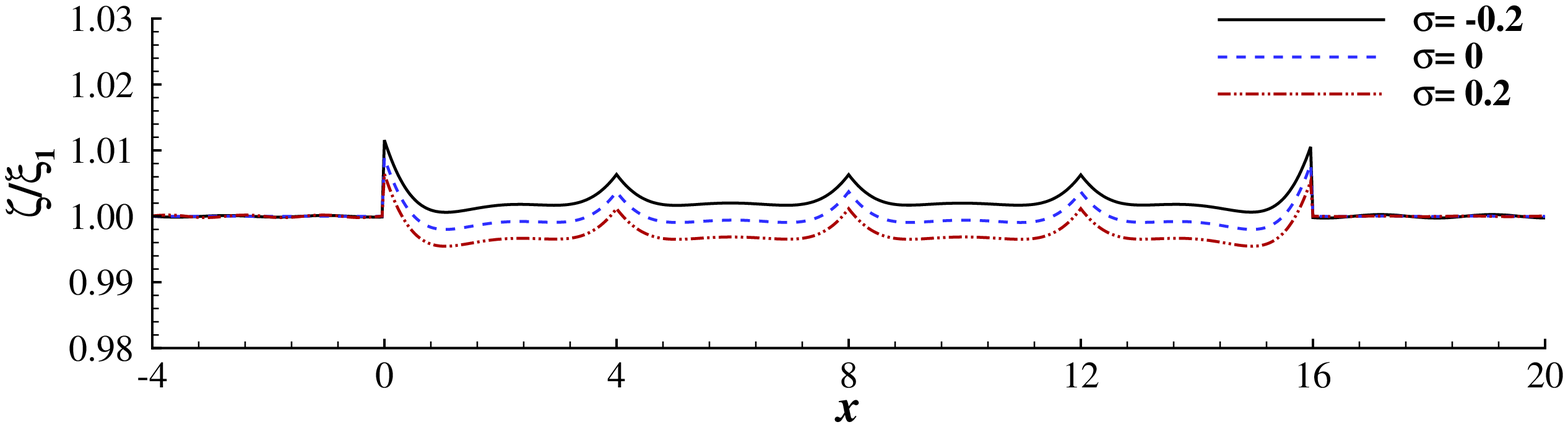}\\
		\makebox[0.9\textwidth][l]{($b$)}
		\includegraphics[width=\textwidth]{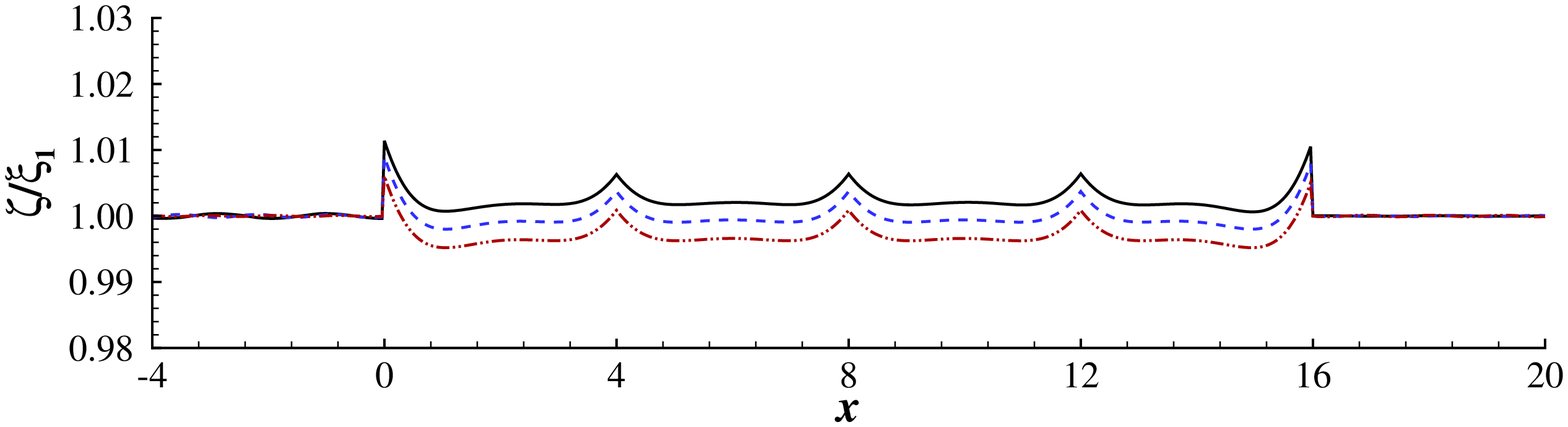}
		\caption{Displacement amplitude on surface affected by different density distributions in a ($a$) 4-layer fluid, ($b$) 8-layer fluid}
		\label{Fig.dis_omeg_sigma}
 \end{figure}

\begin{figure}[H]
		\centering
		\makebox[0.9\textwidth][l]{($a$)}
		\includegraphics[width=\textwidth]{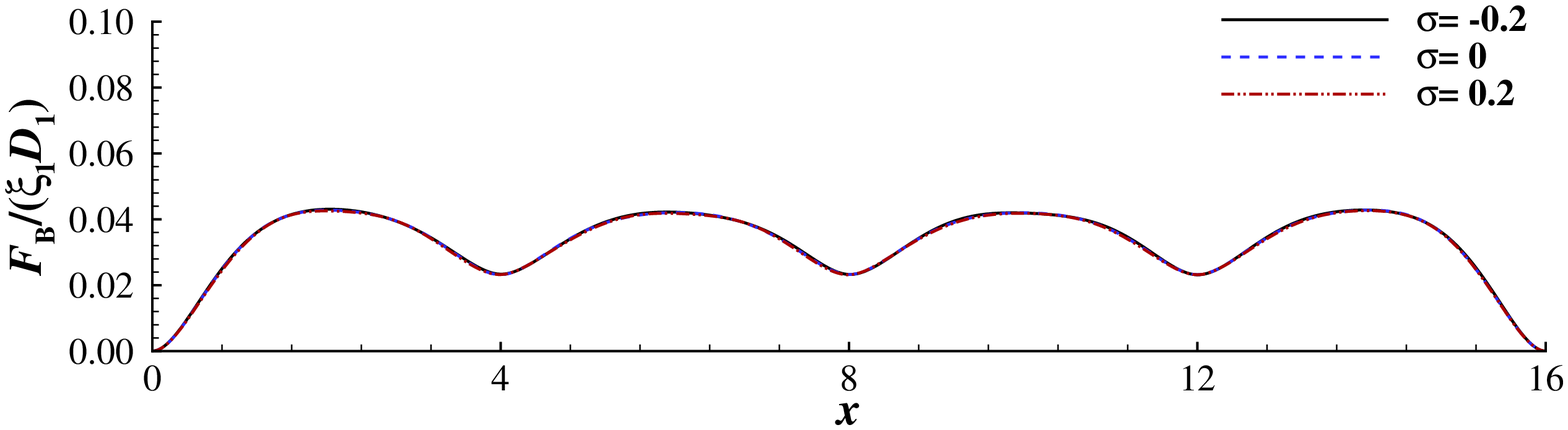}\\
		\makebox[0.9\textwidth][l]{($b$)}
		\includegraphics[width=\textwidth]{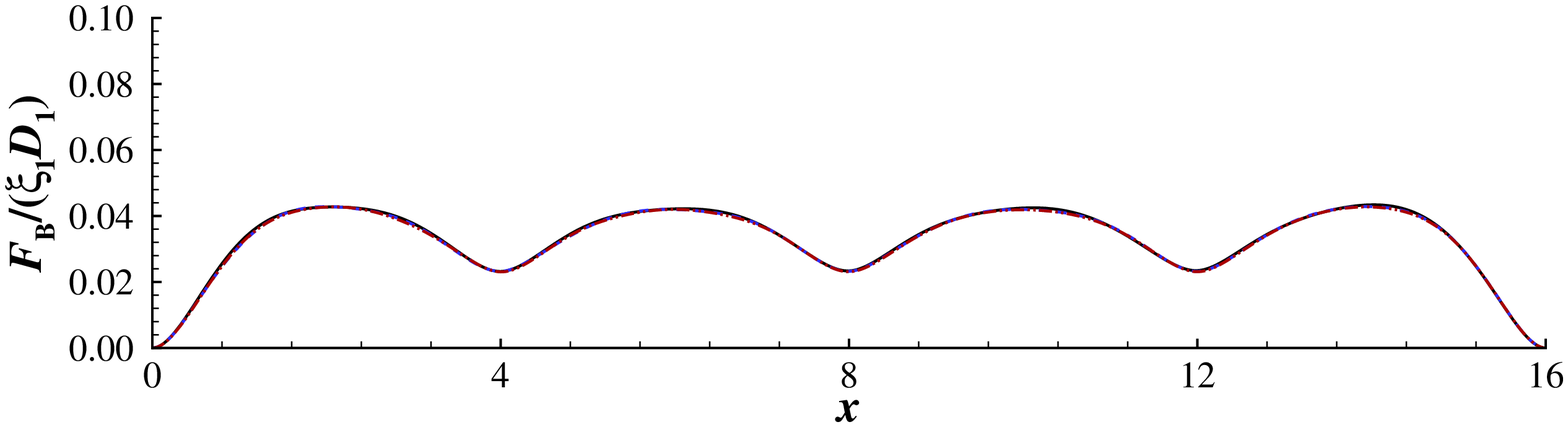}
		\caption{Amplitude of bending moment affected by different density distributions in a ($a$) 4-layer fluid, ($b$) 8-layer fluid}
		\label{Fig.bdm_omeg_sigma}
 \end{figure}

\begin{figure}[H]
		\centering
		\makebox[0.9\textwidth][l]{($a$)}
		\includegraphics[width=\textwidth]{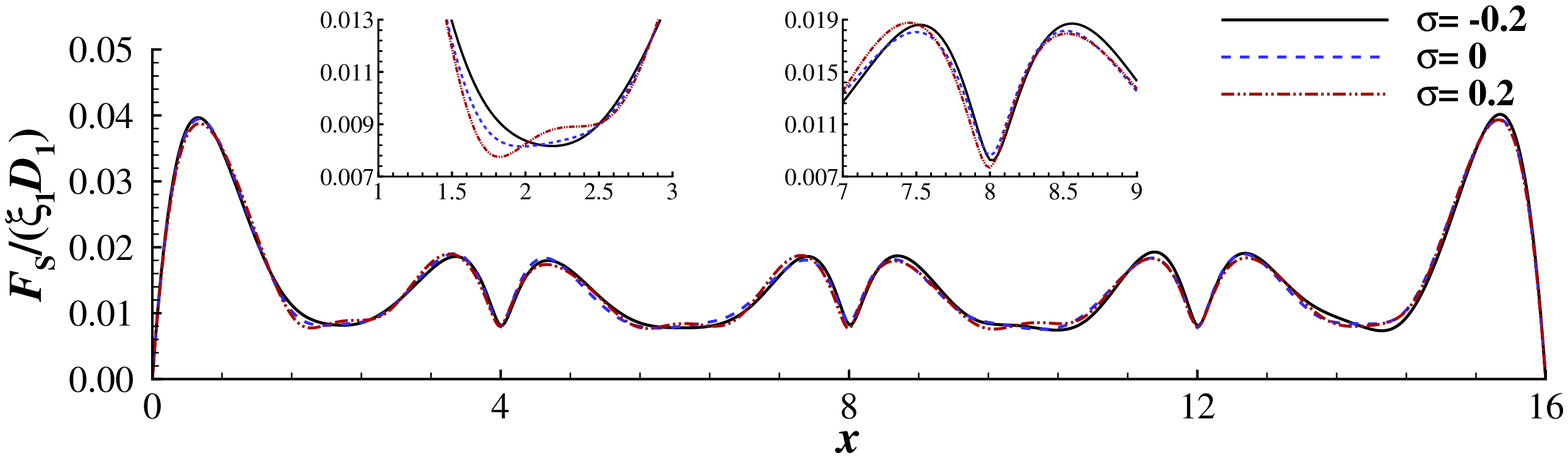}\\
		\makebox[0.9\textwidth][l]{($b$)}
		\includegraphics[width=\textwidth]{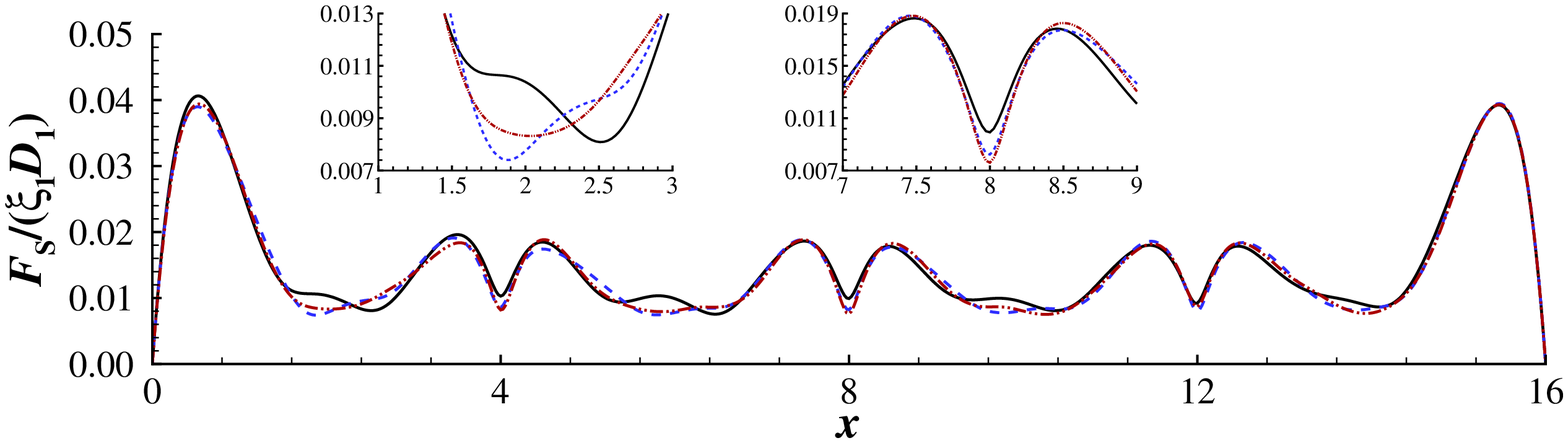}
		\caption{Amplitude of shear force affected by different density distributions in a ($a$) 4-layer fluid, ($b$) 8-layer fluid}
		\label{Fig.shf_omeg_sigma}
 \end{figure}

 Figures \ref{Fig.dis_omeg_sigma}--\ref{Fig.shf_omeg_sigma} show the calculation results for the amplitudes of surface displacement, bending moment and shear force. For the surface displacement, it is noted that the curves in the area $0<x<16$, namely the amplitude of the plate deflection, {falls sightly and stepwise with }$\sigma$ increasing from $-0.2$ to 0.2. But the curve profiles show no difference between the 4-layer fluid and the 8-layer one. For the amplitude of bending moment, it is totally unaffected either by $\sigma$ or by the various stratifications. While for the shear force, a conspicuous variation is exhibited for the values along the whole structure, especially at the middle area of every single plate and the neighborhoods nearby the connections, which has been plotted in the subgraphs. {This phenomenon becomes more intense in the 8-layer fluid, which implies the fluid stratification will generate essential impact on the inner shear forces of the floating elastic plates. Investigations on a more refined fluid stratification are necessarily important.}

\section{\label{Sec.5.con}Conclusions}

 We formulate the scattering problem of flexural--gravity waves across a boundary between the domains covered by elastic plates, and subsequently promote the formulations to the investigation on a generalized extension, i.e.\ multiple elastic plates floating on a stratified fluid with multiple layers, where the numbers of the plates as well as the layers can be arbitrary. The significance of the structural combination of multiple elastic plates is to study the cracked ice floes or multi-module VLFSs, which are usually featured by inhomogeneous local properties. By modeling the fluid as a multi-layer stratified system, the real ocean circumstance with a continuous density distribution can be well approximated by calculating a very precise hierarchy. The method of matched eigenfunction expansions is employed to solve the potential functions within the frame of linear potential theory, in which the inner product technique is widely used to deal with the matching relations. A conventional manner of making an inner product may not have the qualification to lead to a well-conditioned system of equations, especially for multiple matching boundaries underneath elastic plates. Considering that the completeness of the vertical eigenfunction of flexural--gravity waves has never been proved before, we introduce the vertical eigenfunctions of free-surface waves to make an inner product. Under this definition, an orthogonal relation with an explicit differential term is figured out for the two kind of vertical eigenfunctions, which is also adequate to be promoted to the generalized situation for a multi-layer fluid.

 A high evaluating efficiency and a great rate of convergence are assured for the calculations, where an energy conservation relation is used for {verification}. The correctness is also checked via comparing with a previous analytical method. Investigations on the hydrodynamic features of multi-module VLFSs and the impact of stratified fluids are carried out. For a floating structure {with a fixed total length,} we find the 4-plate division will provide a better performance in deformations and internal forces in comparison with a single-plate model and a 2-plate one. For the stratified fluid, we try to introduce a parabolic curve to approximate the pycnocline in the upper area, numbers of discrete points on which are drawn out as representatives for the density of every layer. The numerical results show the shear force will be greatly affected by different stratifications, which is a essential distinction comparing to a uniform fluid.

\begin{acknowledgments}
 This research was sponsored by the National Natural Science Foundation of China under Grant No.\ 11472166 and the National Basic Research Program of China under Grant No.\ 2014CB046203.
\end{acknowledgments}

\bibliographystyle{Refs_citation_order}
\bibliography{bibliography}

% \end{document}

\newpage


\end{document}